\documentclass[%
 aip,
 amsmath,amssymb,
 reprint,%
]{revtex4-1}
\usepackage{macros}

\usepackage{graphicx}% Include figure files
\usepackage{dcolumn}% Align table columns on decimal point
\usepackage{bm}% bold math
\usepackage[mathlines]{lineno}% Enable numbering of text and display math
\usepackage[utf8]{inputenc}
\usepackage[T1]{fontenc}
\usepackage{mathptmx}
\usepackage{etoolbox}

\usepackage{xcolor}

\makeatletter
\def\@email#1#2{%
 \endgroup
 \patchcmd{\titleblock@produce}
  {\frontmatter@RRAPformat}
  {\frontmatter@RRAPformat{\produce@RRAP{*#1\href{mailto:#2}{#2}}}\frontmatter@RRAPformat}
  {}{}
}%
\makeatother

\graphicspath{{Figures}}

\begin{document}

% \preprint{AIP/123-QED}

\title[]{Accurate deep learning-based filtering for chaotic dynamics by identifying instabilities without an ensemble}

% Force line breaks with \\
\author{Marc Bocquet}
\email{marc.bocquet@enpc.fr.}
\homepage{https://cerea.enpc.fr/HomePages/bocquet/}

\author{Alban Farchi}
\altaffiliation{Now at the European Centre for Medium-Range Weather Forecasts (ECMWF), Bonn, Germany}

\author{Tobias S. Finn}
\author{Charlotte Durand}
\author{Sibo Cheng}

\affiliation{CEREA, \'{E}cole des Ponts and EDF R\&D, \^Ile-de-France, France}

\author{Yumeng Chen}
\author{Ivo Pasmans}

\affiliation{Department of Meteorology and National Centre for Earth Observation, University of Reading, Earley Gate, PO
  Box 243, Reading, RG6 6BB, United Kingdom}

\author{Alberto Carrassi}

\affiliation{Department of Physics and Astronomy, University of Bologna, Viale Carlo Berti Pichat, 6/2, Bologna, 40127, Italy}

\date{\today}% It is always \today, today,

\begin{abstract}
  We investigate the ability to discover data assimilation (DA) schemes meant for chaotic dynamics with deep learning.
  The focus is on learning the analysis step of sequential DA, from state trajectories and their observations, using a
  simple residual convolutional neural network, while assuming the dynamics to be known.  Experiments are performed with
  the Lorenz 96 dynamics, which display spatiotemporal chaos and for which solid benchmarks for DA performance exist.
  The accuracy of the states obtained from the learned analysis approaches that of the best possibly tuned ensemble
  Kalman filter, and is far better than that of variational DA alternatives.  Critically, this can be achieved while
  propagating even just a single state in the forecast step. We investigate the reason for achieving ensemble filtering
  accuracy without an ensemble. We diagnose that the analysis scheme actually identifies key dynamical perturbations,
  mildly aligned with the unstable subspace, from the forecast state alone, without any ensemble-based covariances
  representation.  This reveals that the analysis scheme has learned some multiplicative ergodic theorem associated to
  the DA process seen as a non-autonomous random dynamical system.
\end{abstract}

\maketitle

\begin{quotation}
  Data assimilation (DA) estimates the state of dynamical systems from sparse and noisy observations, and is used
  worldwide in numerical weather prediction centers. Accurate DA demands the representation of the time-dependent errors
  in this state estimate, usually achieved through the propagation of an ensemble of states. Using deep learning, we
  discover the update step of DA applied to chaotic dynamics. We show that a simple convolutional neural network (CNN)
  can learn DA, reaching an accuracy as good as that of ensemble-based DA. Crucially, the CNN can achieve this best
  accuracy with single state forecasts. This is explained by the CNN's ability to identify local space patterns from
  this one state, which are used to assess the errors in the analysis. This suggests building a new class of efficient
  deep learning-based ensemble-free DA algorithms.
\end{quotation}

\section{Introduction}
\label{sec:introduction}

\subsection{Context and problem}

In a simplified but quintessential framework, the goal of data assimilation (DA), and in particular filtering
algorithms, is to accurately estimate states $\xtk\in\R^{\Nx}$, where ``t'' stands for truth, at equally spaced times
$\tau_k$ for $k=0,\ldots,K$ along a trajectory of a dynamical system. Hence, they are related by
\begin{subequations}
\label{eq:data-assimilation-operator}
  \begin{equation}
  \label{eq:forecast-operator}
  \xtkp = {\mathcal M}\(\xtk\),
\end{equation}
where $\mathcal M$ is the resolvent over $\tau_{k+1}-\tau_k$ of known autonomous, i.e. time-independent, dynamics.
Such goal is achieved from the knowledge of the dynamics $\mathcal M$ and of observation vectors $\by_k \in \R^{\Nyk}$
obtained from the non-accessible states $\xtk$ via observation operators $\Hc_k$,
and perturbed by a white-in-time Gaussian noise $\errok$ of mean $\bzero$ and covariance matrix $\bR_k$:
\begin{equation}
  \label{eq:observation-operator}
  \by_k = \Hc_k(\xtk)+\errok, \qquad \errok \sim N(\bzero, \bR_k).
\end{equation}
\end{subequations}
Applied to chaotic hence dynamically unstable dynamics, sequential (in time) algorithms must be
used.\citep{asch2016,carrassi2018} They alternate an \emph{analysis} step which, from the newly acquired observation vector
$\by_k$ in Eq.~\eqref{eq:observation-operator} and the current estimate of the state $\xfk$, provides an updated optimal
estimate of the state $\xak$ called the analysis. The subsequent state estimate $\xfkp$ stems from the \emph{forecast} step
which relies on Eq.~\eqref{eq:forecast-operator}.  The estimates in both steps can either be deterministic or
probabilistic, often leveraging an ensemble in the latter case. Such sequential DA is widely used in numerical weather
prediction (NWP), and in many areas of climate sciences,\citep{carrassi2018} as a suite of both research and operational
tools.

Classical DA methods are classified into (i) variational methods, such as 3D-Var and 4D-Var, (ii) ensemble-based
statistical methods, such as the ensemble Kalman filter (EnKF), and (iii) ensemble variational methods which inherit the
assets of the two previous categories.\citep{asch2016} On the one hand, variational methods account for the nonlinearity
of models (dynamical model and observation operators), leveraging nonlinear optimization techniques.  Ensemble-based
methods on the other hand can capture the \emph{errors of the day}, i.e. time-dependent error statistics, via an
ensemble meant to diagnose sample error statistics. Those are key properties that drive the performance of these DA
methods in mildly nonlinear chaotic models. For low-order, chaotic dynamics such as the celebrated Lorenz 96 (L96)
model,\citep{lorenz1998} the EnKF significantly outperforms 3D-Var, or a moderately-long window 4D-Var in terms of
accuracy, owing to its dynamical representation of the errors. This has been emphasized and illustrated in twin
experiments.\citep{bocquet2013a} In fact, current implementations of 4D-Var in NWP centers incorporate a forecast
ensemble so as to capture the errors-of-the-day.\citep{raynaud2012,bonavita2012} However, in high-dimensional models,
these ensemble-based error statistics must necessarily be regularized using techniques known as localization and
possibly inflation.\citep{evensen2009b} With a focus on the time-dependent error statistics of sequential DA, it has
been conjectured \citep{carrassi2008d,palatella2013a} then proven \citep{gurumoorthy2017,bocquet2017a,crisan2023} that
for linear dynamics and when localization is unnecessary, the forecast and analysis error covariance matrices of the
EnKF are confined to the unstable-neutral subspace, denoted as $\uns$ from now on, of the dynamics. This subspace is
spanned by the covariant Lyapunov vectors associated to non-negative Lyapunov exponents.\citep{legras1996} It is
precisely when the ensemble size is smaller than the dimension of this subspace that localization is required to avoid
divergence of the EnKF.  Deviating from linear dynamics turns those exact results into approximations, for which these
findings were nonetheless numerically confirmed.\citep{bocquet2017b,chen2021,carrassi2022}

This paper focuses on methodological DA and on what can be discovered from deep learning (DL) techniques to improve
state-of-the-art DA schemes such as those mentioned above. Hence, we hereby give a brief account on the recent
introduction of DL techniques for DA applied to chaotic dynamics.\citep{cheng2023}

It was first proposed to learn DA analysis through DL from the data produced by existing DA
schemes. \citep{harter2012,cintra2018} One can alternatively replace the solver of a 4D-Var over a long DA window by a
DL operator that would learn the outcome of the 4D-Var cost function
minimization.\citep{fablet2021,frerix2021,lafon2023,filoche2023,keller2024} However, the latter approaches do not
consider cycling sequential DA, the focus of the present paper.  A systematic, formal Bayesian view on the use of DL in
the critical components of sequential DA has been proposed \citep{boudier2023} and called \emph{data assimilation
network} (DAN). In the present paper, a simplified variant of this DAN concept is used.  As far as ensemble and
Kalman-related DA methods are concerned, it has been proposed to learn their Kalman gain, \citep{hoang1994,hoang1998} or
parameters thereof, possibly relying on an auto-differentiable implementation of the
(En)KF. \citep{haarnoja2016,chen2022,luk2024} As a step further, it was also proposed to learn the full analysis operator
using (self-)supervision.\citep{mccabe2021,boudier2023} Finally, bypassing the need for dynamical models and DA schemes
altogether, DL-based \emph{end-to-end} methods aim at estimating states of the system from the observations
only,\citep{mcnally2024b,vaughan2024} yet so far with a focus on the feasibility of such endeavor.

\subsection{Objectives}
In this paper, the forecast model in Eq.~\eqref{eq:forecast-operator} is assumed to be known so as to avoid intricate
interactions when learning the DA operators and the dynamics simultaneously.

Our objective is to learn the analysis operator of a sequential DA scheme meant for chaotic dynamics from a long
trajectory of the dynamical system and the associated set of noisy, possibly sparse observations. Hence, it stands out
from past studies that exploited DL to learn the dynamics, possibly using
DA. \citep{bocquet2019b,brajard2020,bocquet2020,brajard2021,liu2022,wang2024} The resulting DL-based analysis operator
will be referred to as $\dan$, while the full resulting DA scheme will be called DAN.

We will first explain how to learn such analysis operator from DL.  It will then be shown numerically that $\dan$ can
surprisingly perform as accurately as a well optimized EnKF, even \emph{without using an ensemble} which strongly
contrasts with the common beliefs in methodological DA.  To interpret this result, we will show using innovative
concepts based on a Taylor expansion of the learned $\dan$, that $\dan$ directly discovers and utilizes a fine knowledge
of the dynamics, as opposed to agnostic classical DA.  The nature of these dynamical structures learned through
$\dan$ will then be discussed and interpreted.

\section{Experimental setup}
\label{sec:setup}

With the goal to learn an analysis operator $\dan$ as a key step of a filtering DA scheme for chaotic dynamics, we
build a twin experiment within the framework offered by Eqs.~\eqref{eq:data-assimilation-operator}.

\subsection{Analysis operator and its neural network representation}

Let us define a filtering DA scheme, based on an analysis and forecast ensemble.  The $i$-th
members of the analysis and forecast ensembles at time $\tau_k$ are noted $\xaki$ and $\xfki$, respectively.  Denoting
$\Se=1,\ldots,\Ne$, the corresponding analysis and forecast ensembles are $\Eak = \left\{\xaki\right\}_{i\in\Se} \subset
\EEe$, $\Efk = \left\{\xfki\right\}_{i\in\Se} \subset \EEe$, respectively, where $\EEe = \R^{\Ne\times\Nx}$. The initial
ensemble $\Efo$ is obtained from perturbing a random state on the attractor of the dynamics.

The analysis step of the DA scheme is given by the (incremental) analysis operator $\dan$, which depends on a set of
neural network weights and biases, a vector $\btheta$,
\begin{subequations}
  \label{eq:analysis}
  \begin{equation}
    \label{eq:dan}
    \Eak = \Efk+\dan\(\Efk, \bH_k\T\bRi_k\innk\),
  \end{equation}
  where $\innk$, the innovation at time $\tau_k$, is defined by
  \begin{equation}
    \innk \eqdef \by_k - \Hc_k\(\xfkm\), \quad \xfkm \eqdef \frac{1}{\Ne}\sum_{i\in\Se} \xfki.
  \end{equation}
\end{subequations}
$\bH_k$ is the tangent linear operator of $\Hc_k$ but any arbitrary injective operator from $\R^{\Nyk}$ to $\R^{\Nx}$
could be chosen instead.  The DA forecast step propagates the analysis ensemble, member-wise:
\begin{equation}
  \label{eq:forecast}
  \Efkp = {\mathcal M}\(\Eak\).
\end{equation}

\begin{figure}[t]
  \includegraphics[width=0.48\textwidth]{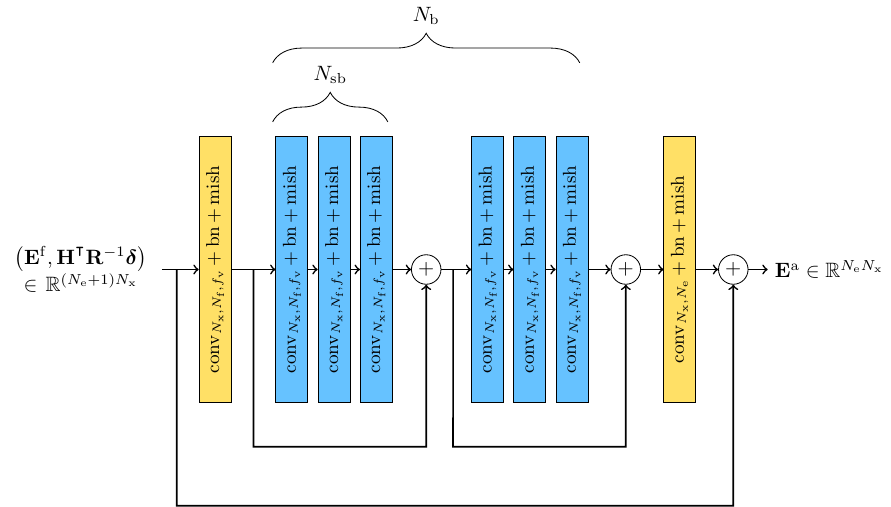}
  \caption{ \label{fig:nn} Architecture of the residual convolutional network, where $\Nb=2$,
    $\Nsb=3$. $\mathrm{conv}_{N_1,N_2,f}$ is a generic one-dimensional convolutional layer of dimension $N_1$, with
    $N_2$ filters of kernel size $f$. See text for more details.}
\end{figure}

We choose $\dan$ to have a simple residual convolutional neural network (CNN) architecture.  A schematic of the
CNN architecture is displayed in Fig.~\ref{fig:nn}.
It begins with an initial convolution that takes $\Ne+1$ channels as inputs and, with $\Nf$ filters, outputs $\Nf$
channels.  This initial layer is followed by $\Nb$ residual blocks. Each one of these blocks is a succession of $\Nsb$
sub-blocks.  Each subblock is made of: (i) a convolutional layer with $\Nf$ channels as inputs, which has $\Nf$ filters
and a kernel size $\fov$ for each of its filter, (ii) a batch normalization layer, and (iii) an activation function
chosen to be mish. \citep{misra2019} The CNN ends with a final convolutional layer that takes $\Nf$ channels as inputs
and, with $\Ne$ filters, outputs $\Ne$ channels.  The kernel size of the initial and final channels is $\fov$.  Hence,
the internal state of the CNN consists of $\Nf$ copies of the latent space which we simply choose to be isomorphic to
the state space $\R^{\Nx}$. Furthermore, the encoder and decoder from state space to latent space and back are chosen to
coincide with the identity. We have also tested a depth-separable architecture for this residual CNN, with a number of
parameters roughly divided by $3$, yielding an accuracy almost as good but longer training times.  Note that the fundamental
results reported in this paper are agnostic to the details of the architecture: this CNN is a mere functional tool to
learn an optimal $\dan$.

\subsection{Training scheme}

Towards efficiently learning an optimal $\dan$, we consider $\Nr$ such DA runs, based on as many independent concurrent
trajectories of the dynamics and as many sequences of observation vectors.  Hence, the DA runs are specified by
$\Eakr=\left\{\xakri\right\}_{i\in\Se}$ and $\Efkr$ for $r=1,\ldots,\Nr$ and, being iterates through $\Nc$ cycles, they
depend on $\btheta$ except for the set of initial conditions $\Efor$.  In order to learn an optimal $\dan$, a loss
function is defined:
\begin{equation}
  \label{eq:loss}
        {\mathcal L}(\btheta) = \sum_{r=1}^{\Nr} \sum_{k=1}^{\Nc} \left\| \xtkr-\xakrm(\btheta) \right\|^2,
        \quad \xakrm \eqdef \frac{1}{\Ne}\sum_{i\in\Se} \xakri,
\end{equation}
where $\left\|\cdot\right\|$ is the Euclidean norm.  Its formulation is based on supervised learning, although
self-supervised learning\citep{bocquet2019b,brajard2020,mccabe2021} could have been used instead; it is nonetheless more challenging
and rather unrelated to the goals of this paper.  This loss matches the analysis ensemble mean trajectory
with the true trajectory. The Adam stochastic gradient descent optimization technique \citep{kingma2015} is used to
minimize it.  To avoid the risks of exploding gradients, the inefficiency of vanishing gradients, and huge memory
requirements, when computing gradients of Eq.~\eqref{eq:loss}, the truncated backpropagation through time technique
\citep{tang2018,aicher2020} is used; it splits the trajectories in the dataset into chunks of $\Niter$ cycles.

It must be pointed out that a successful sequential DA process, when seen as a dynamical system, is
stable. \citep{carrassi2008a,carrassi2008b} Hence, after a rough starting phase in the training, the learned $\dan$
should yield a numerically stable prediction-assimilation dynamical system.  In particular, this is likely to avoid
exploding gradients. By contrast, the task of learning a DL emulator of the dynamics over many consecutive time steps often
fails because of the unstable nature of the chaotic dynamics.

The $\Nr$ trajectories are dispatched into a training and validation dataset with a $90\%-10\%$ ratio. Overfitting is
prevented by an early stopping of the minimization based on the validation score, tantamount to regularization.\citep{goodfellow2016}
Moreover, the testing dataset stems from an independently generated trajectory, long enough to yield converged
statistics. In the test stage, the DAN
scheme is used within a twin experiment using the trajectories and resulting observations of the testing dataset. Its
performance is assessed from a single scalar score using the time-averaged root mean square error (aRMSE) of the
analysis against the truth:
\begin{equation}
\mathrm{aRMSE}=\frac{1}{K\sqrt{\Nx}}\sum_{k=1}^{K}  \left\|\xtk-\xakm\right\|,
\end{equation}
which, in a cycled DA context, is a reliable indicator of the overall performance of the scheme, whatever its purpose.

\section{Numerical results}
\label{sec:results}

The $\dan$ operator is trained on the L96 model,\citep{lorenz1998} and the results will be interpreted and discussed in
the context of this model.  L96 is a one-dimensional model defined over a periodic band of latitude of the Earth
atmosphere.  Its ordinary differential equations read
\begin{equation}
  \frac{{\mathrm{d}}x_n}{{\mathrm{d}}t} = (x_{n+1}-x_{n-2})x_{n-1}-x_n+F ,
\end{equation}
with $x_{N_x}=x_0$, $x_{-1} = x_{N_x-1}$, $x_{-2}=x_{N_x-2}$, $F=8$, and $\Nx=40$ in the basic configuration.  The model
has a Lyapunov time of $0.60$. It has $13$ positive exponents and, being continuous-in-time and autonomous, it has one
zero Lyapunov exponent. Hence, the dimension of its unstable-neutral subspace $\uns$, is $\Nu=14$.

\subsection{Hyperparameters sensitivity analysis}
\label{sec:sensitivity-study}

We first carry out a large set of trainings to assess the sensitivity of $\dan$'s performance to its hyperparameters.
We choose $\Niter=16$, without any significant gain beyond this value while the numerical cost increases due to a deeper
backpropagation.  We first assume the model to be fully observed with $\Hc_k=\Ix$, the identity matrix in $\R^{\Nx}$, and the
observations to be affected by a white-in-time unbiased Gaussian noise of covariance matrix $\bR_k=\Ix$, for all time
steps.  This configuration is the most widely used to benchmark new DA schemes with L96.  The ensemble size $\Ne$ and the number of
filter $\Nf$ were selected in a set ranging between $1$ and $40$.  The number $\Nb$ of residual blocks in the CNN and
number of subblocks $\Nsb$ in each residual block were both chosen in the set $\llbracket 1, 6 \rrbracket$.  Because 
L96 has short-range correlations in space, we choose a kernel size of $\fov=5$, even though the CNN receptive
field is much larger.

\subsection{First results and robustness}

The training dataset size per epoch scales linearly with $\Nr$, which is chosen to be $2^{18}$ and further discussed in
the supplementary material.  The subsequent test DA runs with the trained $\dan$ are actually all stable in time,
yielding an aRMSE significantly below $1$, as expected if DA has any skill over the mere observations.  Unsurprisingly,
we found that the larger the hyperparameters $\Nf,\Nb,\Nsb$, the smaller the test aRMSEs of the resulting DANs, but that
$\Nf=40$, $\Nb=5$, and $\Nsb=5$ offer a good compromise for accuracy versus training cost and CNN size. This will be the
reference configuration, which has about $2\times 10^5$ trainable parameters.

One obvious essential drawback of learning $\dan$ is its \emph{non-universality}. Specifically, $\dan$ depends on the
observation setup used in the training dataset.  This is a critical research path for end-to-end DA. Although not the
aim of the present paper, we nonetheless checked the performance of the trained $\dan$, with $\Hc_k=\Ix$ and
$\bR_k=\sigy^2\Ix$ with $\sigy=1$, in test DA runs with similar observations but generated with $\sigy$ taking value in
between $0.1$ and $3$. Yet, in all test runs, DAN remains robust with slightly degraded aRMSEs for $\sigy<1$ but
aRMSEs at least as good for $\sigy>1$, compared to a well-tuned EnKF.  Well-tuned EnKF always refers here to an EnKF
with an ensemble large enough such that localization is unnecessary and relying on the EnKF-N
\citep{bocquet2015b,raanes2019a} to optimally counteract residual sampling errors such that inflation is unnecessary.

Testing non-trivial $\Hc_k$, we also learned a single $\dan$ from observation networks whose density $\Ny/\Nx$ is
randomly and uniformly chosen in the interval $[0, 1]$ at each $\tau_k$, and $\sigy=1$. This DAN was then tested on
several DA runs, each one with a constant in time observation density $\Ny/\Nx$ taking value in the interval
$[0.2,1]$. In this configuration, $\dan$ performs almost as well or better than well-tuned EnKFs for $0.35<\Ny/\Nx<0.65$
and is suboptimal (compared to the EnKF) but still stable outside of this range. These results already pleasantly
suggest that these DL-based DA schemes may remain valid well beyond the specifications of observation operators from
which $\dan$ was learned.  Plots of these experiments and further discussion are provided in the supplementary material.

\subsection{One state forecast}
Using the reference configuration but with an ensemble size $\Ne$ taking values in the set $\llbracket 1, 40
\rrbracket$, test aRMSEs fluctuate in between $0.19$ and $0.20$. By contrast, a sizable ensemble is, as we recalled in
Sec.~\ref{sec:introduction}, one of the key reason for the success of the EnKF.  For quantitative comparison, we checked
that 3D-Var scores an aRMSE of $0.40$, that the best linear filter, i.e. $\dan$ learned without activation functions scores
$0.384$, whereas well-tuned EnKFs with $\Ne=20$ and $\Ne=40$ score $0.191$ and $0.179$, respectively.  Note that the
reference $\dan$ but with $\Nf=100$ yields an aRMSE of $0.185$, closer to the best EnKF with $\Ne=\Nx=40$, showing that
further improvements are possible even though not the focus of this paper. These key aRMSE scores are arranged in a table in
the supplementary material.

However, the pivotal remark is that a single state forecast, $\Ne=1$ in $\dan$, is as efficient as using a large
ensemble. Furthermore, the need for localization and inflation is completely obviated.  We have checked that this is
obtained concurrently to a feature collapse in $\dan$,\citep{vanamersfoort2022} i.e. all channels' last layer feature
maps converge to the same state. It is likely that a better local minimum of the loss could be obtained with complex
encoder and decoder \citep{peyron2021} and infusing diversity in the CNN through Monte Carlo dropouts,\citep{mccabe2021}
so as to obtain an $\dan$ leveraging the ensemble. Nonetheless, the local minimum reached in our trainings, yield an
accuracy with $\Ne=1$ worthy of a well-tuned EnKF. That is why we shall concentrate in the following on interpreting
this astonishing result for which we shall use, especially in Sec.~\ref{sec:interpretation}, dynamical systems theory.

Therefore, the analysis operator is hereafter learned in the reference configuration but with $\Ne=1$.

\section{Interpretation}
\label{sec:interpretation}

In this section, we focus on the remarkable finding that a learned DA method with a single state $\Ne=1$ forecast achieves
performance on par with a well-tuned EnKF.  We wish to understand the reason for this performance by investigating 
what $\dan$ learns. To that end, an innovative expansion of $\dan$ in terms of more familiar DA operators is carried
out.

\subsection{Operator expansion of $\dan$}

Towards this goal, we look for a classical Kalman update \citep{kalman1960,ghil1991} that would be a good
match to $\dan$ seen as a mathematical map, at least for small analysis increments.  The first diagnostic is the mean
anomaly generated by $\dan$, i.e. how much $\dan(\bx,\bzero)$ deviates from $\bzero$ on average. It should be small
since a vanishing innovation $\innov_k$ should not yield any state update.  Hence, we define the time-dependent
normalized scalar anomalies
\begin{equation}
  \label{eq:bias}
  b_k = \frac{1}{\sqrt{\Nx}}\left\| \dan(\bx_k,\bzero)\right\|,
\end{equation}
along with the associated mean bias $b$ and the standard deviation $s$ of $b_k$ in time.

Next, expanding with respect to the innovation, the following functional form for $\dan$ is assumed:
\begin{equation}
  \label{eq:K-approx}
  \dan(\bx, \bH\T\bRi\innov) \approx \bK(\bx)\cdot\innov,
\end{equation}
owing to the fact that no state update is needed when the innovation vanishes, and only keeping the leading order term
in $\innov$.  This is an Ansatz of $\dan$ where $\bK(\bx)\in\R^{\Nx\times\Ny}$ is meant to stand as a Kalman gain
surrogate.  By contrast, with the propagation of a single state, classical sequential DA methods would typically
resemble 3D-Var, and the gain would not depend on the forecast state (the first input variable of $\dan$).

Interestingly, we also learned a simplified $\hdan$ replacing Eq.~\eqref{eq:dan} with $\Eak =
\Efk+\hdan\(\bH_k\T\bRi_k\innk\)$, whereby losing $\dan$'s ability to extract information from $\Efk$, similarly to
3D-Var.  This yields an aRMSE of $0.382$ in test DA runs, unsurprisingly close to the $0.40$ of our 3D-Var. Hence,
learning an optimal constant-in-time $\bK$ of an (En)KF,\citep{luk2024} a configuration subsumed by this specific
$\hdan$, is significantly suboptimal in this context.

\subsection{Identifying the operators in this expansion}

Once $\dan$ has been obtained from training, and considering a fixed forecast state $\bx$ at a given time step, a
large set of innovations $\left\{\innov_j\right\}_{j=1,\ldots,\Np}$ are sampled from the observation error statistics:
$\innov_j \sim N(\bzero,\bR)$.  This yields a set of corresponding incremental updates
$\left\{\ba_j=\dan(\bx,\bH\T\bRi\innov_j)\right\}_{j=1,\ldots,\Np}$.  Since Eq.~\eqref{eq:K-approx} is only an
approximation, $\bK(\bx)$ is estimated with the least squares problem
\begin{equation}
  \label{eq:least-squares}
  \Lc_\bx(\bK) = \sum_{j=1}^{\Np} \left\| \ba_j-\barba-\bK(\bx)\cdot\(\innov_j-\barinnov\)\right\|^2,
\end{equation}
where $\barba = \Np^{-1}\sum_{j=1}^{\Np}\ba_j$ and $\barinnov=\Np^{-1}\sum_{j=1}^{\Np}\innov_j$.

Next, assuming $\bR$ is known, we would like to estimate the analysis error covariance matrix $\Pa$ associated to $\dan$
in the Kalman gain expansion. It depends on $\bx_k$ and hence on $\tau_k$.  Within the \emph{best linear unbiased
estimator} framework, $\bK$ is related to $\Pa$ through $\bK=\Pa\bH\T\bRi$, so that from Eq.~\eqref{eq:K-approx}:
\begin{equation}
  \dan(\bx, \bH\T\bRi\innov) \approx \Pa\bH\T\bRi\innov,
\end{equation}
which suggests that an expansion in the second variable $\bzeta \in \R^{\Nx}$ of $\dan$ yields
\begin{equation}
  \label{eq:Pa-approx}
  \dan(\bx, \bzeta) \approx \Pa(\bx)\cdot\bzeta.
\end{equation}
Hence, $\Pa$ can be estimated using Eq.~\eqref{eq:Pa-approx} either from a least squares loss similar to
Eq.~\eqref{eq:least-squares} or from the Jacobian of
 $\dan$ with respect to $\bzeta$ leveraging auto-differentiable DL libraries.

\subsection{What is learned? - Supporting numerical results}

At each $\tau_k$, i.e. over many $\bx_k$ on the forecast model's attractor, it is possible to estimate
$\bK(\bx_k)$ and $\Pa(\bx_k)$ from the expansion of $\dan$.  For the sake of simplicity, $\Hc_k=\Ix$, $\bR_k=\Ix$, in
which case $\Pa(\bx_k)=\bK(\bx_k)$.

The analysis mean bias $b$ and its standard deviation $s$ are first computed over a long L96 $\dan$-based DA run.  We
obtain $b\simeq 5.10^{-3}$ and $s\simeq 10^{-3}$ which are indeed very small compared to the typical aRMSE of an either
DAN or EnKF run, i.e. $0.20$, meaning that the bias of $\dan$ relative to typical updates is roughly $2.5\%$.

\begin{figure}[t]
  \includegraphics[width=0.45\textwidth]{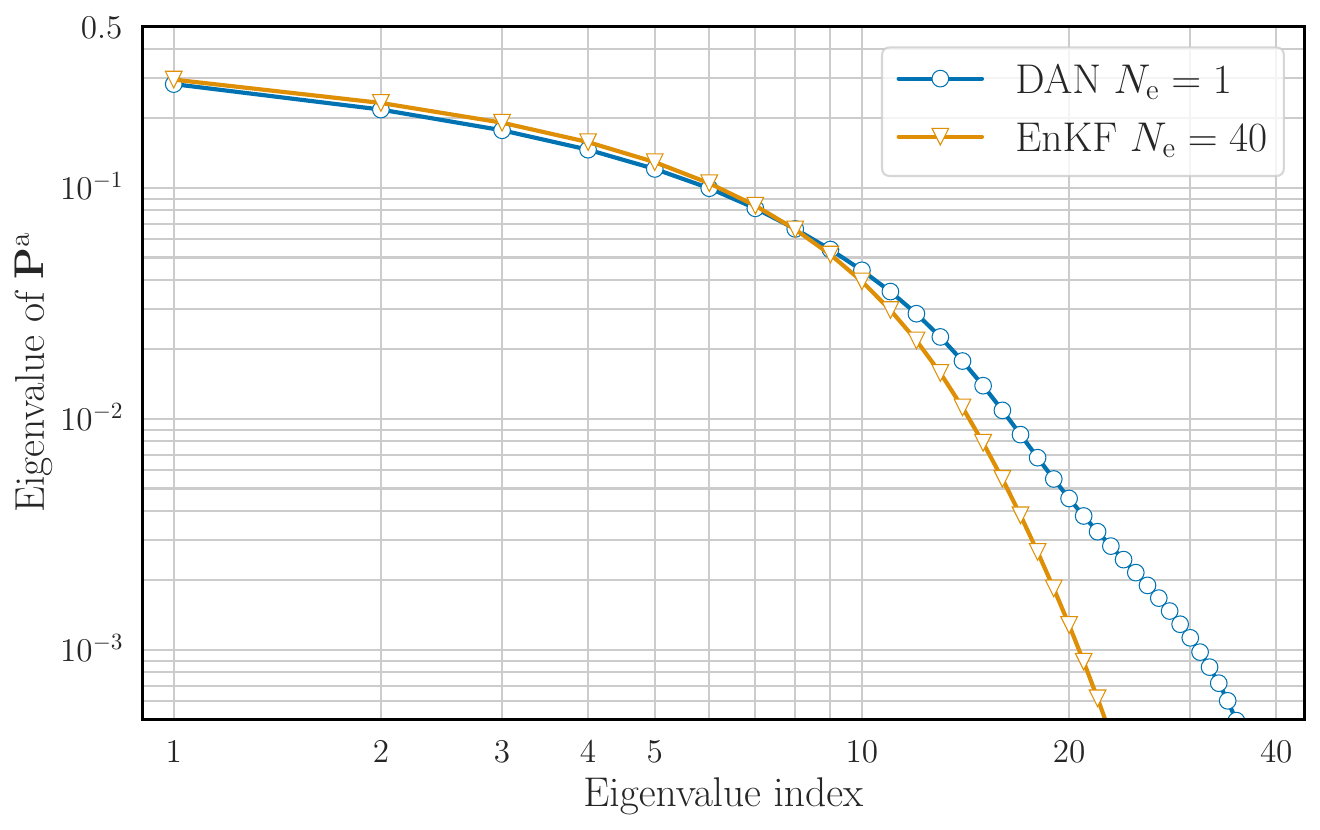}
  \caption{ \label{fig:Pa_spectra_l96} Time-averaged eigenspectra of $\Pa_\mathrm{DAN}$ and $\Pa_\mathrm{EnKF}$.}
\end{figure}

The surrogate $\Pa$, denoted $\Pa_\mathrm{DAN}$ and estimated from Eq.~\eqref{eq:Pa-approx}, is compared to that of a
concurrent well-tuned EnKF with $\Ne=40$, whose analysis error covariance matrix is $\Pa_\mathrm{EnKF}$.
$\Pa_\mathrm{DAN}$ is compared to $\Pa_\mathrm{EnKF}$ using a normalized Bures-Wasserstein distance: \citep{bhatia2019a}
\begin{equation}
  d_\mathrm{BW}(\bU,\bV) = \frac{1}{\Nx} \left[ \tr \left\{\bU + \bV -2 \(\bV\poh\bU\bV\poh\)\poh\right\} \right]\poh,
\end{equation}
where $\bU$ and $\bV$ are two semi-definite symmetric matrices.  This metric is expected to smoothly account for the
unmatched principal axes of $\bU$ and $\bV$, but also their associated variances (eigenspectra).  The time-averaged
$d_\mathrm{BW}$ distance between $\Pa_\mathrm{DAN}$ and $\Pa_\mathrm{EnKF}$ is $0.013$ whereas it is $0.048$ between
$\Pa_\mathrm{DAN}$ and $(0.40)^2\Ix$, which approximates $\Pa$ of a well-tuned 3D-Var. The
time-averaged eigenspectra of $\Pa_\mathrm{DAN}$ and $\Pa_\mathrm{EnKF}$ are plotted in
Fig.~\ref{fig:Pa_spectra_l96}.  They are remarkably close to each other for the first $10$ modes.  Beyond these modes
the $\dan$ operator is likely to selectively apply some (multiplicative) inflation, as one would expect from such stable
DA runs.

\begin{figure}[t]
  \includegraphics[width=0.45\textwidth]{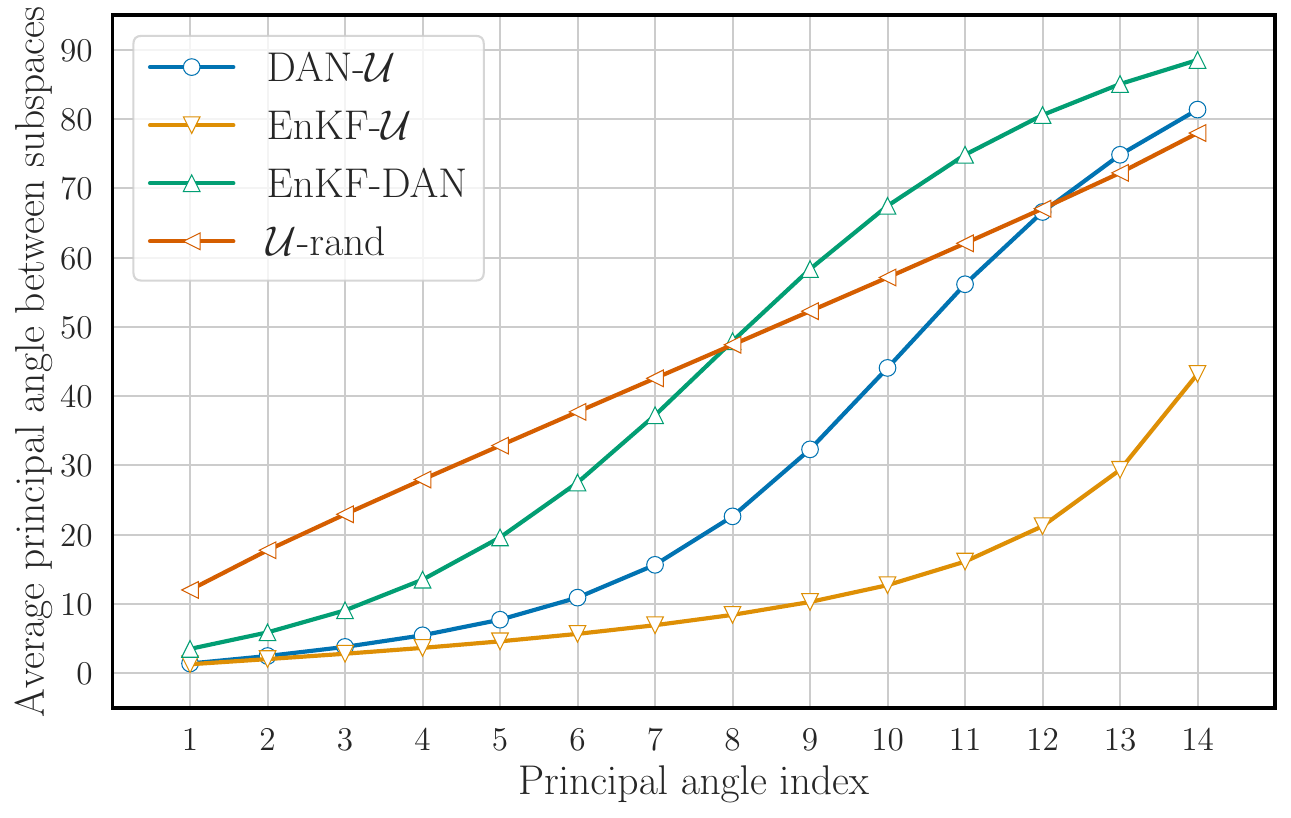}
  \caption{ \label{fig:pang_l96} Time-averaged principal angles (in degrees) formed by the subspaces spanned by the
    $\Nu=14$ dominant directions of $\Pa_\mathrm{DAN}$ versus $\uns$, $\Pa_\mathrm{EnKF}$ versus $\uns$,
    $\Pa_\mathrm{EnKF}$ versus $\Pa_\mathrm{DAN}$, and $\uns$ versus a randomly sampled $\Nu=14$--dimensional subspace.}
\end{figure}

We further compute the principal angles\citep{bocquet2017b} of the vector subspaces generated by the $\Nu=14$ dominant
eigenvalues of $\Pa_\mathrm{DAN}$ and of $\Pa_\mathrm{EnKF}$. They are reported in Fig.~\ref{fig:pang_l96}.  Recall that
$\Nu=14$ is the dimension of the L96 $\uns$. The principal angles are intrinsic to the relative position of these
subspaces; they do not depend on any coordinate system used to parameterize them.  This indicates how close the most
unstable directions of $\Pa_\mathrm{DAN}$ and $\Pa_\mathrm{EnKF}$ are in state space. From Fig.~\ref{fig:pang_l96}, we
observe that the simplex formed by the EnKF is on average the most aligned with $\uns$.\citep{bocquet2017b} The subspace
spanned by the dominant axes of $\Pa_\mathrm{DAN}$ are also well aligned with $\uns$, yet progressively diverges when
incorporating less unstable directions.  For comparison, the principal angles of $\uns$ with an isotropically randomly
sampled $\Nu=14$-dimensional subspace are also plotted in Fig.~\ref{fig:pang_l96}.

\subsection{Main interpretation}
These numerical results indicate that $\dan$ defined through Eq.~\eqref{eq:dan}, depends on the innovation, but also on
the single forecast state when $\Ne=1$. This does not hold for the EnKF incremental update which only indirectly depends
on the forecast state via the ensemble-based forecast error covariances. Hence, without the need for an ensemble, $\dan$
extracts from the forecast state critical pieces of information on the unstable directions, as shown by the principal
angles experiment.

Furthermore, $\dan$ manages to accurately assess the uncertainty attached to these unstable directions as demonstrated
by the spectra of $\Pa_\mathrm{DAN}$. Overall, $\Pa_\mathrm{DAN}$ with $\Ne=1$ is on average very close to
$\Pa_\mathrm{EnKF}$ with $\Ne=40$, for the dominant axes, and it applies some inflation onto the less unstable modes as
seen by comparing their spectra.\citep{bocquet2015b} We conclude that $\dan$ directly learns about the dynamics
features, as opposed to the regression-based, purely statistical, update in the EnKF.

Essentially, for $\dan$, critical pieces of information of the forecast error covariances of the DA run are encoded, and
thus exploitable, in the forecast state alone.  From the multiplicative ergodic theorem,\citep{oseledec1968} we know
that, in autonomous ergodic dynamical systems such as $\mathcal M$, there exists a mapping between each of the system's
states and the corresponding Lyapunov covariant vectors.  Further, if the DA run (the forecast and analysis cycle) is
considered as an ergodic dynamical system of its own,\cite{carrassi2008a} the same theorem guarantees the existence of a
mapping between the forecast state and the analysis error covariance matrix that $\dan$ guesses.  The DA process is not
autonomous because it indirectly depends on the truth trajectory, the observation noise, and the observation operators;
but a generalized variant of the multiplicative ergodic theorem for non-autonomous random dynamics should be
applicable. \citep{arnold1998,chekroun2011,flandoli2021,ghil2023} Hence, we conjecture that $\dan$ must learn such
mapping, together with how to process this information and combine it with the innovation.

\subsection{Locality and scalability}

Next, we have trained $\dan$ on the L96 model using the reference configuration with $\Ne=1$, but with a changing state
space dimension $\Nx$ in between $20$ and $160$. The aRMSEs of well-tuned EnKFs for the changing $\Nx$, and picking
$\Ne=\Nx$, has been computed for comparison. The test DAN aRMSEs shows no significant dependence on $\Nx$ and are all
within $5\%$ of the EnKFs.  Hence, because the performance of $\dan$ with an unchanged architecture and the same number
of parameters is barely affected by increasing $\Nx$, we conjecture that the learned analysis extracts \emph{local}
pieces of information from the forecast state.

If true, the $\dan$ operator learned for DA on an $\Nx=40$ L96 model could be applied directly to an L96 DA run with a
different $\Nx$. Recall that the L96 states exhibit local highs and lows of Rossby-like waves, whose number
scales linearly with $\Nx$.  Thus as long as the spatial extent of those waves is captured by the receptive field of the
CNN, the same layers of $\dan$ with the same weights and biases might be able to handle L96 states of distinct
dimensionality.

\begin{figure}[t]
\includegraphics[width=0.45\textwidth]{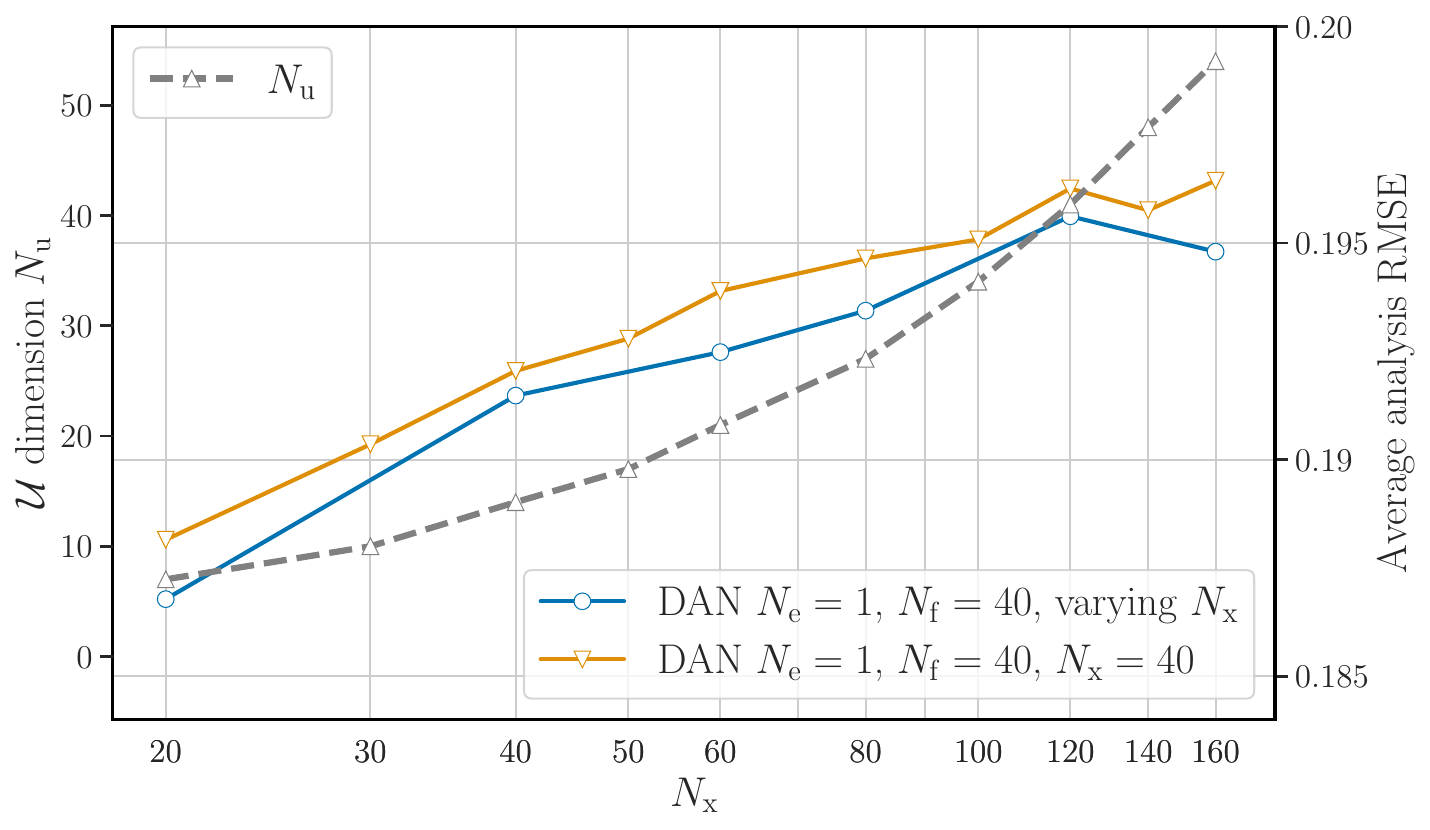}
  \caption{ \label{fig:dan_perf_transfer_from_40_to_Nx_l96} Test aRMSEs (blue and yellow full lines) of $\dan$ operators
    learned from either L96 models with varying $\Nx$, or learned from the $\Nx=40$ L96 model but applied to varying
    $\Nx$ L96 models. The dimension $\Nu$ of $\uns$ (gray dashed line) is much steeper compared to the slowly increasing
    aRMSE curves.}
\end{figure}

To test this hypothesis, we use the same $\dan$ operator (same weights and biases) learned as before with $\Ne=1$,
$\Nx=40$ but apply it now to L96 models with $\Nx$ ranging from $20$ to $160$.  The corresponding aRMSEs are reported in
Fig.~\ref{fig:dan_perf_transfer_from_40_to_Nx_l96}, which shows that these aRMSEs are roughly the same for all $\Nx$
(between 0.188 and 0.197). This demonstrates that this \emph{transdimensional transfer} works surprisingly well.
This strongly supports the fact that $\dan$ extracts local information from the forecast state (of various dimensions in
this experiment), relying on its convolution layers.  It is therefore able to capture where, in phase-space, the error
mass is concentrated. We hypothesize that these localized error structures are related to the \emph{localization} of the
dominant covariant Lyapunov vectors.\citep{pazo2010,vannitsem2016,giggins2019} A proper mathematical definition of such
spatial localization can be found in these references.

\section{Conclusions}

Using the L96 chaotic model we have demonstrated that a learned DL-based analysis $\dan$, key part of a sequential DA
(often referred to as a filtering scheme) can be almost as accurate as the best possibly tuned EnKF, the benchmark for
ensemble filtering methods in this model.  More importantly, this learned DA scheme does not require any ensemble and
can equally well rely on a single state forecast.  Therefore, $\dan$ appears to be able to retrieve local patterns,
representative of unstable and uncertain modes, from the forecast state alone.  We believe that this is fundamentally
made possible by some multiplicative ergodic theorem applied to sequential DA seen as a non-autonomous random dynamical
system driven by time-dependent true dynamics and observation operators, and white-in-time observation errors.

To make sure our conclusions were not entirely bound to the L96 model, we carried out a large number of similar
experiments on the well-known chaotic Kuramoto-Sivashinski model.\citep{kuramoto1976,sivashinsky1977} They all confirm
and support these conclusions.

What is achieved by $\dan$ resonates with the \emph{parametric EnKF}, \citep{pannekoucke2016,pannekoucke2018} which
encodes the errors of the day in a couple of dynamical ancillary fields, preventing the use of an ensemble. Amazingly,
our learned $\dan$ is even more radical and extracts that information from the state itself.

Taking a step back, we learned from DL that an accurate and efficient DA analysis operator could capture the dynamical
error without an ensemble, leveraging model-specific information. This promotes a rethinking of the popular sequential
DA schemes for chaotic dynamics.

\section*{Supplementary material}

See the supplementary material for further details on the evaluation of the $\dan$-based DA method, as reported in
Sec.~\ref{sec:results}.  Specifically, a table of the key aRMSE scores mentioned Sec.~\ref{sec:results} is provided, as
well as plots of the aRMSE curves related to the sensitivity experiments mentioned in Sec.~\ref{sec:results}.

\begin{acknowledgments}
The authors are grateful to the Editor J\"urgen Kurths and an anonymous Reviewer for their comments and suggestions on
the original version of the manuscript.  The authors acknowledge the support of the project SASIP (grant
no. $G-24-66154$) funded by Schmidt Sciences -- a philanthropic initiative that seeks to improve societal outcomes
through the development of emerging science and technologies. CEREA is a member of Institut Pierre-Simon Laplace (IPSL).
\end{acknowledgments}

\section*{Data Availability Statement}
No datasets were used in this article.

%% \nocite{*}
\bibliography{references}% Produces the bibliography via BibTeX.

\begin{thebibliography}{65}%
\makeatletter
\providecommand \@ifxundefined [1]{%
 \@ifx{#1\undefined}
}%
\providecommand \@ifnum [1]{%
 \ifnum #1\expandafter \@firstoftwo
 \else \expandafter \@secondoftwo
 \fi
}%
\providecommand \@ifx [1]{%
 \ifx #1\expandafter \@firstoftwo
 \else \expandafter \@secondoftwo
 \fi
}%
\providecommand \natexlab [1]{#1}%
\providecommand \enquote  [1]{``#1''}%
\providecommand \bibnamefont  [1]{#1}%
\providecommand \bibfnamefont [1]{#1}%
\providecommand \citenamefont [1]{#1}%
\providecommand \href@noop [0]{\@secondoftwo}%
\providecommand \href [0]{\begingroup \@sanitize@url \@href}%
\providecommand \@href[1]{\@@startlink{#1}\@@href}%
\providecommand \@@href[1]{\endgroup#1\@@endlink}%
\providecommand \@sanitize@url [0]{\catcode `\\12\catcode `\$12\catcode
  `\&12\catcode `\#12\catcode `\^12\catcode `\_12\catcode `\%12\relax}%
\providecommand \@@startlink[1]{}%
\providecommand \@@endlink[0]{}%
\providecommand \url  [0]{\begingroup\@sanitize@url \@url }%
\providecommand \@url [1]{\endgroup\@href {#1}{\urlprefix }}%
\providecommand \urlprefix  [0]{URL }%
\providecommand \Eprint [0]{\href }%
\providecommand \doibase [0]{http://dx.doi.org/}%
\providecommand \selectlanguage [0]{\@gobble}%
\providecommand \bibinfo  [0]{\@secondoftwo}%
\providecommand \bibfield  [0]{\@secondoftwo}%
\providecommand \translation [1]{[#1]}%
\providecommand \BibitemOpen [0]{}%
\providecommand \bibitemStop [0]{}%
\providecommand \bibitemNoStop [0]{.\EOS\space}%
\providecommand \EOS [0]{\spacefactor3000\relax}%
\providecommand \BibitemShut  [1]{\csname bibitem#1\endcsname}%
\let\auto@bib@innerbib\@empty
%</preamble>
\bibitem [{\citenamefont {Asch}, \citenamefont {Bocquet},\ and\ \citenamefont
  {Nodet}(2016)}]{asch2016}%
  \BibitemOpen
  \bibfield  {author} {\bibinfo {author} {\bibfnamefont {M.}~\bibnamefont
  {Asch}}, \bibinfo {author} {\bibfnamefont {M.}~\bibnamefont {Bocquet}}, \
  and\ \bibinfo {author} {\bibfnamefont {M.}~\bibnamefont {Nodet}},\ }\href
  {\doibase 10.1137/1.9781611974546} {\emph {\bibinfo {title} {Data
  {A}ssimilation: {M}ethods, {A}lgorithms, and {A}pplications}}},\ Fundamentals
  of Algorithms\ (\bibinfo  {publisher} {SIAM, Philadelphia},\ \bibinfo {year}
  {2016})\ p.\ \bibinfo {pages} {324}\BibitemShut {NoStop}%
\bibitem [{\citenamefont {Carrassi}\ \emph {et~al.}(2018)\citenamefont
  {Carrassi}, \citenamefont {Bocquet}, \citenamefont {Bertino},\ and\
  \citenamefont {Evensen}}]{carrassi2018}%
  \BibitemOpen
  \bibfield  {author} {\bibinfo {author} {\bibfnamefont {A.}~\bibnamefont
  {Carrassi}}, \bibinfo {author} {\bibfnamefont {M.}~\bibnamefont {Bocquet}},
  \bibinfo {author} {\bibfnamefont {L.}~\bibnamefont {Bertino}}, \ and\
  \bibinfo {author} {\bibfnamefont {G.}~\bibnamefont {Evensen}},\ }\bibfield
  {title} {\enquote {\bibinfo {title} {Data assimilation in the geosciences: An
  overview on methods, issues, and perspectives},}\ }\href {\doibase
  10.1002/wcc.535} {\bibfield  {journal} {\bibinfo  {journal} {WIREs Climate
  Change}\ }\textbf {\bibinfo {volume} {9}},\ \bibinfo {pages} {e535} (\bibinfo
  {year} {2018})}\BibitemShut {NoStop}%
\bibitem [{\citenamefont {Lorenz}\ and\ \citenamefont
  {Emanuel}(1998)}]{lorenz1998}%
  \BibitemOpen
  \bibfield  {author} {\bibinfo {author} {\bibfnamefont {E.~N.}\ \bibnamefont
  {Lorenz}}\ and\ \bibinfo {author} {\bibfnamefont {K.~A.}\ \bibnamefont
  {Emanuel}},\ }\bibfield  {title} {\enquote {\bibinfo {title} {Optimal sites
  for supplementary weather observations: simulation with a small model},}\
  }\href {\doibase 10.1175/1520-0469(1998)055<0399:OSFSWO>2.0.CO;2} {\bibfield
  {journal} {\bibinfo  {journal} {J. Atmos. Sci.}\ }\textbf {\bibinfo {volume}
  {55}},\ \bibinfo {pages} {399--414} (\bibinfo {year} {1998})}\BibitemShut
  {NoStop}%
\bibitem [{\citenamefont {Bocquet}\ and\ \citenamefont
  {Sakov}(2013)}]{bocquet2013a}%
  \BibitemOpen
  \bibfield  {author} {\bibinfo {author} {\bibfnamefont {M.}~\bibnamefont
  {Bocquet}}\ and\ \bibinfo {author} {\bibfnamefont {P.}~\bibnamefont
  {Sakov}},\ }\bibfield  {title} {\enquote {\bibinfo {title} {Joint state and
  parameter estimation with an iterative ensemble {K}alman smoother},}\ }\href
  {\doibase 10.5194/npg-20-803-2013} {\bibfield  {journal} {\bibinfo  {journal}
  {Nonlin. Processes Geophys.}\ }\textbf {\bibinfo {volume} {20}},\ \bibinfo
  {pages} {803--818} (\bibinfo {year} {2013})}\BibitemShut {NoStop}%
\bibitem [{\citenamefont {Raynaud}, \citenamefont {Berre},\ and\ \citenamefont
  {Desroziers}(2012)}]{raynaud2012}%
  \BibitemOpen
  \bibfield  {author} {\bibinfo {author} {\bibfnamefont {L.}~\bibnamefont
  {Raynaud}}, \bibinfo {author} {\bibfnamefont {L.}~\bibnamefont {Berre}}, \
  and\ \bibinfo {author} {\bibfnamefont {G.}~\bibnamefont {Desroziers}},\
  }\bibfield  {title} {\enquote {\bibinfo {title} {Accounting for model error
  in the {Météo-France} ensemble data assimilation system},}\ }\href
  {\doibase 10.1002/qj.906} {\bibfield  {journal} {\bibinfo  {journal} {Q. J.
  R. Meteorol. Soc.}\ }\textbf {\bibinfo {volume} {138}},\ \bibinfo {pages}
  {249--262} (\bibinfo {year} {2012})}\BibitemShut {NoStop}%
\bibitem [{\citenamefont {Bonavita}, \citenamefont {Isaksen},\ and\
  \citenamefont {H{\'o}lm}(2012)}]{bonavita2012}%
  \BibitemOpen
  \bibfield  {author} {\bibinfo {author} {\bibfnamefont {M.}~\bibnamefont
  {Bonavita}}, \bibinfo {author} {\bibfnamefont {L.}~\bibnamefont {Isaksen}}, \
  and\ \bibinfo {author} {\bibfnamefont {E.}~\bibnamefont {H{\'o}lm}},\
  }\bibfield  {title} {\enquote {\bibinfo {title} {On the use of {EDA}
  background error variances in the {ECMWF} {4D-Var}},}\ }\href {\doibase
  10.1002/qj.1899} {\bibfield  {journal} {\bibinfo  {journal} {Q. J. R.
  Meteorol. Soc.}\ }\textbf {\bibinfo {volume} {138}},\ \bibinfo {pages}
  {1540--1559} (\bibinfo {year} {2012})}\BibitemShut {NoStop}%
\bibitem [{\citenamefont {Evensen}(2009)}]{evensen2009b}%
  \BibitemOpen
  \bibfield  {author} {\bibinfo {author} {\bibfnamefont {G.}~\bibnamefont
  {Evensen}},\ }\href {\doibase 10.1007/978-3-642-03711-5} {\emph {\bibinfo
  {title} {{D}ata {A}ssimilation: {T}he {E}nsemble {K}alman {F}ilter}}},\
  \bibinfo {edition} {2nd}\ ed.\ (\bibinfo  {publisher} {Springer-Verlag Berlin
  Heildelberg},\ \bibinfo {year} {2009})\ p.\ \bibinfo {pages}
  {307}\BibitemShut {NoStop}%
\bibitem [{\citenamefont {Carrassi}\ \emph
  {et~al.}(2008{\natexlab{a}})\citenamefont {Carrassi}, \citenamefont
  {Vannitsem}, \citenamefont {Zupanski}, ,\ and\ \citenamefont
  {Zupanski}}]{carrassi2008d}%
  \BibitemOpen
  \bibfield  {author} {\bibinfo {author} {\bibfnamefont {A.}~\bibnamefont
  {Carrassi}}, \bibinfo {author} {\bibfnamefont {S.}~\bibnamefont {Vannitsem}},
  \bibinfo {author} {\bibfnamefont {D.}~\bibnamefont {Zupanski}}, , \ and\
  \bibinfo {author} {\bibfnamefont {M.}~\bibnamefont {Zupanski}},\ }\bibfield
  {title} {\enquote {\bibinfo {title} {The maximum likelihood ensemble filter
  performances in chaotic systems},}\ }\href {\doibase
  10.1111/j.1600-0870.2009.00408.x} {\bibfield  {journal} {\bibinfo  {journal}
  {Tellus A}\ }\textbf {\bibinfo {volume} {61}},\ \bibinfo {pages} {587--600}
  (\bibinfo {year} {2008}{\natexlab{a}})}\BibitemShut {NoStop}%
\bibitem [{\citenamefont {Palatella}, \citenamefont {Carrassi},\ and\
  \citenamefont {Trevisan}(2013)}]{palatella2013a}%
  \BibitemOpen
  \bibfield  {author} {\bibinfo {author} {\bibfnamefont {L.}~\bibnamefont
  {Palatella}}, \bibinfo {author} {\bibfnamefont {A.}~\bibnamefont {Carrassi}},
  \ and\ \bibinfo {author} {\bibfnamefont {A.}~\bibnamefont {Trevisan}},\
  }\bibfield  {title} {\enquote {\bibinfo {title} {Lyapunov vectors and
  assimilation in the unstable subspace: theory and applications},}\ }\href
  {\doibase 10.1088/1751-8113/46/25/254020} {\bibfield  {journal} {\bibinfo
  {journal} {J. Phys. A: Math. Theor.}\ }\textbf {\bibinfo {volume} {46}},\
  \bibinfo {pages} {254020} (\bibinfo {year} {2013})}\BibitemShut {NoStop}%
\bibitem [{\citenamefont {Gurumoorthy}\ \emph {et~al.}(2017)\citenamefont
  {Gurumoorthy}, \citenamefont {Grudzien}, \citenamefont {Apte}, \citenamefont
  {Carrassi},\ and\ \citenamefont {Jones}}]{gurumoorthy2017}%
  \BibitemOpen
  \bibfield  {author} {\bibinfo {author} {\bibfnamefont {K.~S.}\ \bibnamefont
  {Gurumoorthy}}, \bibinfo {author} {\bibfnamefont {C.}~\bibnamefont
  {Grudzien}}, \bibinfo {author} {\bibfnamefont {A.}~\bibnamefont {Apte}},
  \bibinfo {author} {\bibfnamefont {A.}~\bibnamefont {Carrassi}}, \ and\
  \bibinfo {author} {\bibfnamefont {C.~K. R.~T.}\ \bibnamefont {Jones}},\
  }\bibfield  {title} {\enquote {\bibinfo {title} {Rank deficiency of {K}alman
  error covariance matrices in linear time-varying system with deterministic
  evolution},}\ }\href {\doibase 10.1137/15M1025839} {\bibfield  {journal}
  {\bibinfo  {journal} {SIAM J. Control Optim.}\ }\textbf {\bibinfo {volume}
  {55}},\ \bibinfo {pages} {741--759} (\bibinfo {year} {2017})}\BibitemShut
  {NoStop}%
\bibitem [{\citenamefont {Bocquet}\ \emph {et~al.}(2017)\citenamefont
  {Bocquet}, \citenamefont {Gurumoorthy}, \citenamefont {Apte}, \citenamefont
  {Carrassi}, \citenamefont {Grudzien},\ and\ \citenamefont
  {Jones}}]{bocquet2017a}%
  \BibitemOpen
  \bibfield  {author} {\bibinfo {author} {\bibfnamefont {M.}~\bibnamefont
  {Bocquet}}, \bibinfo {author} {\bibfnamefont {K.~S.}\ \bibnamefont
  {Gurumoorthy}}, \bibinfo {author} {\bibfnamefont {A.}~\bibnamefont {Apte}},
  \bibinfo {author} {\bibfnamefont {A.}~\bibnamefont {Carrassi}}, \bibinfo
  {author} {\bibfnamefont {C.}~\bibnamefont {Grudzien}}, \ and\ \bibinfo
  {author} {\bibfnamefont {C.~K. R.~T.}\ \bibnamefont {Jones}},\ }\bibfield
  {title} {\enquote {\bibinfo {title} {Degenerate {K}alman filter error
  covariances and their convergence onto the unstable subspace},}\ }\href
  {\doibase 10.1137/16M1068712} {\bibfield  {journal} {\bibinfo  {journal}
  {SIAM/ASA J. Uncertain. Quantif.}\ }\textbf {\bibinfo {volume} {5}},\
  \bibinfo {pages} {304--333} (\bibinfo {year} {2017})}\BibitemShut {NoStop}%
\bibitem [{\citenamefont {Crisan}\ and\ \citenamefont
  {Ghil}(2023)}]{crisan2023}%
  \BibitemOpen
  \bibfield  {author} {\bibinfo {author} {\bibfnamefont {D.}~\bibnamefont
  {Crisan}}\ and\ \bibinfo {author} {\bibfnamefont {M.}~\bibnamefont {Ghil}},\
  }\bibfield  {title} {\enquote {\bibinfo {title} {Asymptotic behavior of the
  forecast--assimilation process with unstable dynamics},}\ }\href {\doibase
  10.1063/5.0105590} {\bibfield  {journal} {\bibinfo  {journal} {Chaos}\
  }\textbf {\bibinfo {volume} {33}} (\bibinfo {year} {2023}),\
  10.1063/5.0105590}\BibitemShut {NoStop}%
\bibitem [{\citenamefont {Legras}\ and\ \citenamefont
  {Vautard}(1996)}]{legras1996}%
  \BibitemOpen
  \bibfield  {author} {\bibinfo {author} {\bibfnamefont {B.}~\bibnamefont
  {Legras}}\ and\ \bibinfo {author} {\bibfnamefont {R.}~\bibnamefont
  {Vautard}},\ }\bibfield  {title} {\enquote {\bibinfo {title} {A guide to
  lyapunov vectors},}\ }in\ \href@noop {} {\emph {\bibinfo {booktitle} {ECMWF
  Workshop on Predictability}}}\ (\bibinfo  {publisher} {ECMWF},\ \bibinfo
  {address} {Reading, United-Kingdom},\ \bibinfo {year} {1996})\ pp.\ \bibinfo
  {pages} {135--146}\BibitemShut {NoStop}%
\bibitem [{\citenamefont {Bocquet}\ and\ \citenamefont
  {Carrassi}(2017)}]{bocquet2017b}%
  \BibitemOpen
  \bibfield  {author} {\bibinfo {author} {\bibfnamefont {M.}~\bibnamefont
  {Bocquet}}\ and\ \bibinfo {author} {\bibfnamefont {A.}~\bibnamefont
  {Carrassi}},\ }\bibfield  {title} {\enquote {\bibinfo {title}
  {Four-dimensional ensemble variational data assimilation and the unstable
  subspace},}\ }\href {\doibase 10.1080/16000870.2017.1304504} {\bibfield
  {journal} {\bibinfo  {journal} {Tellus A}\ }\textbf {\bibinfo {volume}
  {69}},\ \bibinfo {pages} {1304504} (\bibinfo {year} {2017})}\BibitemShut
  {NoStop}%
\bibitem [{\citenamefont {Chen}, \citenamefont {Carrassi},\ and\ \citenamefont
  {Lucarini}(2021)}]{chen2021}%
  \BibitemOpen
  \bibfield  {author} {\bibinfo {author} {\bibfnamefont {Y.}~\bibnamefont
  {Chen}}, \bibinfo {author} {\bibfnamefont {A.}~\bibnamefont {Carrassi}}, \
  and\ \bibinfo {author} {\bibfnamefont {V.}~\bibnamefont {Lucarini}},\
  }\bibfield  {title} {\enquote {\bibinfo {title} {Inferring the instability of
  a dynamical system from the skill of data assimilation exercises},}\ }\href
  {\doibase 10.5194/npg-28-633-2021} {\bibfield  {journal} {\bibinfo  {journal}
  {Nonlin. Processes Geophys.}\ ,\ \bibinfo {pages} {633--649}} (\bibinfo
  {year} {2021})}\BibitemShut {NoStop}%
\bibitem [{\citenamefont {Carrassi}\ \emph {et~al.}(2022)\citenamefont
  {Carrassi}, \citenamefont {Bocquet}, \citenamefont {Demaeyer}, \citenamefont
  {Gruzien}, \citenamefont {Raanes},\ and\ \citenamefont
  {Vannitsem}}]{carrassi2022}%
  \BibitemOpen
  \bibfield  {author} {\bibinfo {author} {\bibfnamefont {A.}~\bibnamefont
  {Carrassi}}, \bibinfo {author} {\bibfnamefont {M.}~\bibnamefont {Bocquet}},
  \bibinfo {author} {\bibfnamefont {J.}~\bibnamefont {Demaeyer}}, \bibinfo
  {author} {\bibfnamefont {C.}~\bibnamefont {Gruzien}}, \bibinfo {author}
  {\bibfnamefont {P.~N.}\ \bibnamefont {Raanes}}, \ and\ \bibinfo {author}
  {\bibfnamefont {S.}~\bibnamefont {Vannitsem}},\ }\bibfield  {title} {\enquote
  {\bibinfo {title} {Data assimilation for chaotic dynamics},}\ }in\ \href
  {\doibase 10.1007/978-3-030-77722-7_1} {\emph {\bibinfo {booktitle} {Data
  Assimilation for Atmospheric, Oceanic and Hydrologic Applications (Vol.
  IV)}}},\ \bibinfo {editor} {edited by\ \bibinfo {editor} {\bibfnamefont
  {S.~K.}\ \bibnamefont {P.}}\ and\ \bibinfo {editor} {\bibfnamefont
  {L.}~\bibnamefont {X.}}}\ (\bibinfo  {publisher} {Springer International
  Publishing},\ \bibinfo {address} {Cham},\ \bibinfo {year} {2022})\ pp.\
  \bibinfo {pages} {1--42}\BibitemShut {NoStop}%
\bibitem [{\citenamefont {Cheng}\ \emph {et~al.}(2023)\citenamefont {Cheng},
  \citenamefont {Quilodran-Casas}, \citenamefont {Ouala}, \citenamefont
  {Farchi}, \citenamefont {Liu}, \citenamefont {Tandeo}, \citenamefont
  {Fablet}, \citenamefont {Lucor}, \citenamefont {Iooss}, \citenamefont
  {Brajard}, \citenamefont {Xiao}, \citenamefont {Janjic}, \citenamefont
  {Ding}, \citenamefont {Guo}, \citenamefont {Carrassi}, \citenamefont
  {Bocquet},\ and\ \citenamefont {Arcucci}}]{cheng2023}%
  \BibitemOpen
  \bibfield  {author} {\bibinfo {author} {\bibfnamefont {S.}~\bibnamefont
  {Cheng}}, \bibinfo {author} {\bibfnamefont {C.}~\bibnamefont
  {Quilodran-Casas}}, \bibinfo {author} {\bibfnamefont {S.}~\bibnamefont
  {Ouala}}, \bibinfo {author} {\bibfnamefont {A.}~\bibnamefont {Farchi}},
  \bibinfo {author} {\bibfnamefont {C.}~\bibnamefont {Liu}}, \bibinfo {author}
  {\bibfnamefont {P.}~\bibnamefont {Tandeo}}, \bibinfo {author} {\bibfnamefont
  {R.}~\bibnamefont {Fablet}}, \bibinfo {author} {\bibfnamefont
  {D.}~\bibnamefont {Lucor}}, \bibinfo {author} {\bibfnamefont
  {B.}~\bibnamefont {Iooss}}, \bibinfo {author} {\bibfnamefont
  {J.}~\bibnamefont {Brajard}}, \bibinfo {author} {\bibfnamefont
  {D.}~\bibnamefont {Xiao}}, \bibinfo {author} {\bibfnamefont {T.}~\bibnamefont
  {Janjic}}, \bibinfo {author} {\bibfnamefont {W.}~\bibnamefont {Ding}},
  \bibinfo {author} {\bibfnamefont {Y.}~\bibnamefont {Guo}}, \bibinfo {author}
  {\bibfnamefont {A.}~\bibnamefont {Carrassi}}, \bibinfo {author}
  {\bibfnamefont {M.}~\bibnamefont {Bocquet}}, \ and\ \bibinfo {author}
  {\bibfnamefont {R.}~\bibnamefont {Arcucci}},\ }\bibfield  {title} {\enquote
  {\bibinfo {title} {Machine learning with data assimilation and uncertainty
  quantification for dynamical systems: a review},}\ }\href {\doibase
  10.1109/JAS.2023.123537} {\bibfield  {journal} {\bibinfo  {journal} {IEEE/CAA
  J. Autom. Sin.}\ }\textbf {\bibinfo {volume} {10}},\ \bibinfo {pages}
  {1361--1387} (\bibinfo {year} {2023})}\BibitemShut {NoStop}%
\bibitem [{\citenamefont {H{\"a}rter}\ and\ \citenamefont
  {de~Campos~Velho}(2012)}]{harter2012}%
  \BibitemOpen
  \bibfield  {author} {\bibinfo {author} {\bibfnamefont {T.~P.}\ \bibnamefont
  {H{\"a}rter}}\ and\ \bibinfo {author} {\bibfnamefont {H.~F.}\ \bibnamefont
  {de~Campos~Velho}},\ }\bibfield  {title} {\enquote {\bibinfo {title} {Data
  assimilation procedure by recurrent neural network},}\ }\href {\doibase
  10.1080/19942060.2012.11015417} {\bibfield  {journal} {\bibinfo  {journal}
  {Eng. Appl. Comput. Fluid Mech.}\ }\textbf {\bibinfo {volume} {6}},\ \bibinfo
  {pages} {224--233} (\bibinfo {year} {2012})}\BibitemShut {NoStop}%
\bibitem [{\citenamefont {Cintra}\ and\ \citenamefont
  {de~Campos~Velho}(2018)}]{cintra2018}%
  \BibitemOpen
  \bibfield  {author} {\bibinfo {author} {\bibfnamefont {R.~S.}\ \bibnamefont
  {Cintra}}\ and\ \bibinfo {author} {\bibfnamefont {H.~F.}\ \bibnamefont
  {de~Campos~Velho}},\ }\enquote {\bibinfo {title} {Data assimilation by
  artificial neural networks for an atmospheric general circulation model},}\
  in\ \href {\doibase 10.5772/intechopen.70791} {\emph {\bibinfo {booktitle}
  {Advanced applications for artificial neural networks}}},\ \bibinfo {editor}
  {edited by\ \bibinfo {editor} {\bibfnamefont {A.}~\bibnamefont {ElShahat}}}\
  (\bibinfo  {publisher} {IntechOpen},\ \bibinfo {year} {2018})\ Chap.~\bibinfo
  {chapter} {17}, pp.\ \bibinfo {pages} {265--286}\BibitemShut {NoStop}%
\bibitem [{\citenamefont {Fablet}\ \emph {et~al.}(2021)\citenamefont {Fablet},
  \citenamefont {Chapron}, \citenamefont {Drumetz}, \citenamefont {M{\'e}min},
  \citenamefont {Pannekoucke},\ and\ \citenamefont {Rousseau}}]{fablet2021}%
  \BibitemOpen
  \bibfield  {author} {\bibinfo {author} {\bibfnamefont {R.}~\bibnamefont
  {Fablet}}, \bibinfo {author} {\bibfnamefont {B.}~\bibnamefont {Chapron}},
  \bibinfo {author} {\bibfnamefont {L.}~\bibnamefont {Drumetz}}, \bibinfo
  {author} {\bibfnamefont {E.}~\bibnamefont {M{\'e}min}}, \bibinfo {author}
  {\bibfnamefont {O.}~\bibnamefont {Pannekoucke}}, \ and\ \bibinfo {author}
  {\bibfnamefont {F.}~\bibnamefont {Rousseau}},\ }\bibfield  {title} {\enquote
  {\bibinfo {title} {Learning variational data assimilation models and
  solvers},}\ }\href {\doibase 10.1029/2021MS002572} {\bibfield  {journal}
  {\bibinfo  {journal} {J. Adv. Model. Earth Syst.}\ }\textbf {\bibinfo
  {volume} {13}},\ \bibinfo {pages} {e2021MS002572} (\bibinfo {year}
  {2021})}\BibitemShut {NoStop}%
\bibitem [{\citenamefont {Frerix}\ \emph {et~al.}(2021)\citenamefont {Frerix},
  \citenamefont {Kochkov}, \citenamefont {Smith}, \citenamefont {Cremers},
  \citenamefont {Brenner},\ and\ \citenamefont {Hoyer}}]{frerix2021}%
  \BibitemOpen
  \bibfield  {author} {\bibinfo {author} {\bibfnamefont {T.}~\bibnamefont
  {Frerix}}, \bibinfo {author} {\bibfnamefont {D.}~\bibnamefont {Kochkov}},
  \bibinfo {author} {\bibfnamefont {J.}~\bibnamefont {Smith}}, \bibinfo
  {author} {\bibfnamefont {D.}~\bibnamefont {Cremers}}, \bibinfo {author}
  {\bibfnamefont {M.}~\bibnamefont {Brenner}}, \ and\ \bibinfo {author}
  {\bibfnamefont {S.}~\bibnamefont {Hoyer}},\ }\bibfield  {title} {\enquote
  {\bibinfo {title} {Variational data assimilation with a learned inverse
  observation operator},}\ }in\ \href
  {https://proceedings.mlr.press/v139/frerix21a.html} {\emph {\bibinfo
  {booktitle} {Proceedings of the 38th International Conference on Machine
  Learning}}},\ \bibinfo {series} {Proceedings of Machine Learning Research},
  Vol.\ \bibinfo {volume} {139},\ \bibinfo {editor} {edited by\ \bibinfo
  {editor} {\bibfnamefont {M.}~\bibnamefont {Meila}}\ and\ \bibinfo {editor}
  {\bibfnamefont {T.}~\bibnamefont {Zhang}}}\ (\bibinfo  {publisher} {PMLR},\
  \bibinfo {year} {2021})\ pp.\ \bibinfo {pages} {3449--3458}\BibitemShut
  {NoStop}%
\bibitem [{\citenamefont {Lafon}, \citenamefont {Fablet},\ and\ \citenamefont
  {Naveau}(2023)}]{lafon2023}%
  \BibitemOpen
  \bibfield  {author} {\bibinfo {author} {\bibfnamefont {N.}~\bibnamefont
  {Lafon}}, \bibinfo {author} {\bibfnamefont {R.}~\bibnamefont {Fablet}}, \
  and\ \bibinfo {author} {\bibfnamefont {P.}~\bibnamefont {Naveau}},\
  }\bibfield  {title} {\enquote {\bibinfo {title} {Uncertainty quantification
  when learning dynamical models and solvers with variational methods},}\
  }\href {\doibase 10.1029/2022MS003446} {\bibfield  {journal} {\bibinfo
  {journal} {J. Adv. Model. Earth Syst.}\ }\textbf {\bibinfo {volume} {15}},\
  \bibinfo {pages} {e2022MS003446} (\bibinfo {year} {2023})}\BibitemShut
  {NoStop}%
\bibitem [{\citenamefont {Filoche}\ \emph {et~al.}(2023)\citenamefont
  {Filoche}, \citenamefont {Brajard}, \citenamefont {Charantonis},\ and\
  \citenamefont {B{\'e}r{\'e}ziat}}]{filoche2023}%
  \BibitemOpen
  \bibfield  {author} {\bibinfo {author} {\bibfnamefont {A.}~\bibnamefont
  {Filoche}}, \bibinfo {author} {\bibfnamefont {J.}~\bibnamefont {Brajard}},
  \bibinfo {author} {\bibfnamefont {A.}~\bibnamefont {Charantonis}}, \ and\
  \bibinfo {author} {\bibfnamefont {D.}~\bibnamefont {B{\'e}r{\'e}ziat}},\
  }\bibfield  {title} {\enquote {\bibinfo {title} {Learning {4DVAR} inversion
  directly from observations},}\ }in\ \href {\doibase
  10.1007/978-3-031-36027-5_32} {\emph {\bibinfo {booktitle} {Computational
  Science -- ICCS 2023}}},\ \bibinfo {editor} {edited by\ \bibinfo {editor}
  {\bibfnamefont {J.}~\bibnamefont {Miky{\v{s}}ka}}, \bibinfo {editor}
  {\bibfnamefont {C.}~\bibnamefont {de~Mulatier}}, \bibinfo {editor}
  {\bibfnamefont {M.}~\bibnamefont {Paszynski}}, \bibinfo {editor}
  {\bibfnamefont {V.~V.}\ \bibnamefont {Krzhizhanovskaya}}, \bibinfo {editor}
  {\bibfnamefont {J.~J.}\ \bibnamefont {Dongarra}}, \ and\ \bibinfo {editor}
  {\bibfnamefont {P.~M.}\ \bibnamefont {Sloot}}}\ (\bibinfo  {publisher}
  {Springer Nature Switzerland},\ \bibinfo {address} {Cham},\ \bibinfo {year}
  {2023})\ pp.\ \bibinfo {pages} {414--421}\BibitemShut {NoStop}%
\bibitem [{\citenamefont {Keller}\ and\ \citenamefont
  {Potthast}(2024)}]{keller2024}%
  \BibitemOpen
  \bibfield  {author} {\bibinfo {author} {\bibfnamefont {J.~D.}\ \bibnamefont
  {Keller}}\ and\ \bibinfo {author} {\bibfnamefont {R.}~\bibnamefont
  {Potthast}},\ }\href {\doibase 10.48550/arXiv.2406.00390} {\enquote {\bibinfo
  {title} {{AI}-based data assimilation: {L}earning the functional of analysis
  estimation},}\ } (\bibinfo {year} {2024}),\ \Eprint
  {http://arxiv.org/abs/2406.00390} {arXiv:2406.00390 [physics.ao-ph]}
  \BibitemShut {NoStop}%
\bibitem [{\citenamefont {Boudier}\ \emph {et~al.}(2023)\citenamefont
  {Boudier}, \citenamefont {Fillion}, \citenamefont {Gratton}, \citenamefont
  {G{ü}rol},\ and\ \citenamefont {Zhang}}]{boudier2023}%
  \BibitemOpen
  \bibfield  {author} {\bibinfo {author} {\bibfnamefont {P.}~\bibnamefont
  {Boudier}}, \bibinfo {author} {\bibfnamefont {A.}~\bibnamefont {Fillion}},
  \bibinfo {author} {\bibfnamefont {S.}~\bibnamefont {Gratton}}, \bibinfo
  {author} {\bibfnamefont {S.}~\bibnamefont {G{ü}rol}}, \ and\ \bibinfo
  {author} {\bibfnamefont {S.}~\bibnamefont {Zhang}},\ }\bibfield  {title}
  {\enquote {\bibinfo {title} {Data assimilation networks},}\ }\href {\doibase
  10.1029/2022MS003353} {\bibfield  {journal} {\bibinfo  {journal} {J. Adv.
  Model. Earth Syst.}\ }\textbf {\bibinfo {volume} {15}},\ \bibinfo {pages}
  {e2022MS003353} (\bibinfo {year} {2023})}\BibitemShut {NoStop}%
\bibitem [{\citenamefont {Hoang}, \citenamefont {De~Mey},\ and\ \citenamefont
  {Talagrand}(1994)}]{hoang1994}%
  \BibitemOpen
  \bibfield  {author} {\bibinfo {author} {\bibfnamefont {H.}~\bibnamefont
  {Hoang}}, \bibinfo {author} {\bibfnamefont {P.}~\bibnamefont {De~Mey}}, \
  and\ \bibinfo {author} {\bibfnamefont {O.}~\bibnamefont {Talagrand}},\
  }\bibfield  {title} {\enquote {\bibinfo {title} {A simple adaptive algorithm
  of stochastic approximation type for system parameter and state
  estimation},}\ }in\ \href {\doibase 10.1109/CDC.1994.410863} {\emph {\bibinfo
  {booktitle} {Proceedings of 1994 33rd IEEE Conference on Decision and
  Control}}},\ Vol.~\bibinfo {volume} {1}\ (\bibinfo {year} {1994})\ pp.\
  \bibinfo {pages} {747--752 vol.1}\BibitemShut {NoStop}%
\bibitem [{\citenamefont {Hoang}\ \emph {et~al.}(1998)\citenamefont {Hoang},
  \citenamefont {Baraille}, \citenamefont {Talagrand}, \citenamefont {Carton},\
  and\ \citenamefont {De~Mey}}]{hoang1998}%
  \BibitemOpen
  \bibfield  {author} {\bibinfo {author} {\bibfnamefont {S.}~\bibnamefont
  {Hoang}}, \bibinfo {author} {\bibfnamefont {R.}~\bibnamefont {Baraille}},
  \bibinfo {author} {\bibfnamefont {O.}~\bibnamefont {Talagrand}}, \bibinfo
  {author} {\bibfnamefont {X.}~\bibnamefont {Carton}}, \ and\ \bibinfo {author}
  {\bibfnamefont {P.}~\bibnamefont {De~Mey}},\ }\bibfield  {title} {\enquote
  {\bibinfo {title} {Adaptive filtering: application to satellite data
  assimilation in oceanography},}\ }\href {\doibase
  10.1016/S0377-0265(97)00014-6} {\bibfield  {journal} {\bibinfo  {journal}
  {Dynam. Atmos. Ocean}\ }\textbf {\bibinfo {volume} {27}},\ \bibinfo {pages}
  {257--281} (\bibinfo {year} {1998})}\BibitemShut {NoStop}%
\bibitem [{\citenamefont {Haarnoja}\ \emph {et~al.}(2016)\citenamefont
  {Haarnoja}, \citenamefont {Ajay}, \citenamefont {Levine},\ and\ \citenamefont
  {Abbeel}}]{haarnoja2016}%
  \BibitemOpen
  \bibfield  {author} {\bibinfo {author} {\bibfnamefont {T.}~\bibnamefont
  {Haarnoja}}, \bibinfo {author} {\bibfnamefont {A.}~\bibnamefont {Ajay}},
  \bibinfo {author} {\bibfnamefont {S.}~\bibnamefont {Levine}}, \ and\ \bibinfo
  {author} {\bibfnamefont {P.}~\bibnamefont {Abbeel}},\ }\bibfield  {title}
  {\enquote {\bibinfo {title} {Backprop {KF}: Learning discriminative
  deterministic state estimators},}\ }in\ \href
  {https://proceedings.neurips.cc/paper_files/paper/2016/file/697e382cfd25b07a3e62275d3ee132b3-Paper.pdf}
  {\emph {\bibinfo {booktitle} {Advances in Neural Information Processing
  Systems}}},\ Vol.~\bibinfo {volume} {29},\ \bibinfo {editor} {edited by\
  \bibinfo {editor} {\bibfnamefont {D.}~\bibnamefont {Lee}}, \bibinfo {editor}
  {\bibfnamefont {M.}~\bibnamefont {Sugiyama}}, \bibinfo {editor}
  {\bibfnamefont {U.}~\bibnamefont {Luxburg}}, \bibinfo {editor} {\bibfnamefont
  {I.}~\bibnamefont {Guyon}}, \ and\ \bibinfo {editor} {\bibfnamefont
  {R.}~\bibnamefont {Garnett}}}\ (\bibinfo  {publisher} {Curran Associates,
  Inc.},\ \bibinfo {year} {2016})\BibitemShut {NoStop}%
\bibitem [{\citenamefont {Chen}, \citenamefont {Sanz-Alonso},\ and\
  \citenamefont {Willett}(2022)}]{chen2022}%
  \BibitemOpen
  \bibfield  {author} {\bibinfo {author} {\bibfnamefont {Y.}~\bibnamefont
  {Chen}}, \bibinfo {author} {\bibfnamefont {D.}~\bibnamefont {Sanz-Alonso}}, \
  and\ \bibinfo {author} {\bibfnamefont {R.}~\bibnamefont {Willett}},\
  }\bibfield  {title} {\enquote {\bibinfo {title} {Autodifferentiable ensemble
  kalman filters},}\ }\href {\doibase 10.1137/21M1434477} {\bibfield  {journal}
  {\bibinfo  {journal} {SIAM J. Math. Data Sci.}\ }\textbf {\bibinfo {volume}
  {4}},\ \bibinfo {pages} {801--833} (\bibinfo {year} {2022})}\BibitemShut
  {NoStop}%
\bibitem [{\citenamefont {Luk}\ \emph {et~al.}(2024)\citenamefont {Luk},
  \citenamefont {Bach}, \citenamefont {Baptista},\ and\ \citenamefont
  {Stuart}}]{luk2024}%
  \BibitemOpen
  \bibfield  {author} {\bibinfo {author} {\bibfnamefont {E.}~\bibnamefont
  {Luk}}, \bibinfo {author} {\bibfnamefont {E.}~\bibnamefont {Bach}}, \bibinfo
  {author} {\bibfnamefont {R.}~\bibnamefont {Baptista}}, \ and\ \bibinfo
  {author} {\bibfnamefont {A.}~\bibnamefont {Stuart}},\ }\href {\doibase
  10.48550/arXiv.2406.18066} {\enquote {\bibinfo {title} {Learning optimal
  filters using variational inference},}\ } (\bibinfo {year} {2024}),\ \Eprint
  {http://arxiv.org/abs/2406.18066} {arXiv:2406.18066 [cs.LG]} \BibitemShut
  {NoStop}%
\bibitem [{\citenamefont {McCabe}\ and\ \citenamefont
  {Brown}(2021)}]{mccabe2021}%
  \BibitemOpen
  \bibfield  {author} {\bibinfo {author} {\bibfnamefont {M.}~\bibnamefont
  {McCabe}}\ and\ \bibinfo {author} {\bibfnamefont {J.}~\bibnamefont {Brown}},\
  }\bibfield  {title} {\enquote {\bibinfo {title} {Learning to assimilate in
  chaotic dynamical systems},}\ }in\ \href
  {https://proceedings.neurips.cc/paper_files/paper/2021/file/65cc2c8205a05d7379fa3a6386f710e1-Paper.pdf}
  {\emph {\bibinfo {booktitle} {Advances in Neural Information Processing
  Systems}}},\ Vol.~\bibinfo {volume} {34},\ \bibinfo {editor} {edited by\
  \bibinfo {editor} {\bibfnamefont {M.}~\bibnamefont {Ranzato}}, \bibinfo
  {editor} {\bibfnamefont {A.}~\bibnamefont {Beygelzimer}}, \bibinfo {editor}
  {\bibfnamefont {Y.}~\bibnamefont {Dauphin}}, \bibinfo {editor} {\bibfnamefont
  {P.}~\bibnamefont {Liang}}, \ and\ \bibinfo {editor} {\bibfnamefont {J.~W.}\
  \bibnamefont {Vaughan}}}\ (\bibinfo  {publisher} {Curran Associates, Inc.},\
  \bibinfo {year} {2021})\ pp.\ \bibinfo {pages} {12237--12250}\BibitemShut
  {NoStop}%
\bibitem [{\citenamefont {McNally}\ \emph {et~al.}(2024)\citenamefont
  {McNally}, \citenamefont {Lessig}, \citenamefont {Lean}, \citenamefont
  {Boucher}, \citenamefont {Alexe}, \citenamefont {Pinnington}, \citenamefont
  {Chantry}, \citenamefont {Lang}, \citenamefont {Burrows}, \citenamefont
  {Chrust}, \citenamefont {Pinault}, \citenamefont {Villeneuve}, \citenamefont
  {Bormann},\ and\ \citenamefont {Healy}}]{mcnally2024b}%
  \BibitemOpen
  \bibfield  {author} {\bibinfo {author} {\bibfnamefont {A.}~\bibnamefont
  {McNally}}, \bibinfo {author} {\bibfnamefont {C.}~\bibnamefont {Lessig}},
  \bibinfo {author} {\bibfnamefont {P.}~\bibnamefont {Lean}}, \bibinfo {author}
  {\bibfnamefont {E.}~\bibnamefont {Boucher}}, \bibinfo {author} {\bibfnamefont
  {M.}~\bibnamefont {Alexe}}, \bibinfo {author} {\bibfnamefont
  {E.}~\bibnamefont {Pinnington}}, \bibinfo {author} {\bibfnamefont
  {M.}~\bibnamefont {Chantry}}, \bibinfo {author} {\bibfnamefont
  {S.}~\bibnamefont {Lang}}, \bibinfo {author} {\bibfnamefont {C.}~\bibnamefont
  {Burrows}}, \bibinfo {author} {\bibfnamefont {M.}~\bibnamefont {Chrust}},
  \bibinfo {author} {\bibfnamefont {F.}~\bibnamefont {Pinault}}, \bibinfo
  {author} {\bibfnamefont {E.}~\bibnamefont {Villeneuve}}, \bibinfo {author}
  {\bibfnamefont {N.}~\bibnamefont {Bormann}}, \ and\ \bibinfo {author}
  {\bibfnamefont {S.}~\bibnamefont {Healy}},\ }\href {\doibase
  10.48550/arXiv.2407.15586} {\enquote {\bibinfo {title} {Data driven weather
  forecasts trained and initialised directly from observations},}\ } (\bibinfo
  {year} {2024}),\ \Eprint {http://arxiv.org/abs/2407.15586} {arXiv:2407.15586
  [physics.ao-ph]} \BibitemShut {NoStop}%
\bibitem [{\citenamefont {Vaughan}\ \emph {et~al.}(2024)\citenamefont
  {Vaughan}, \citenamefont {Markou}, \citenamefont {Tebbutt}, \citenamefont
  {Requeima}, \citenamefont {Bruinsma}, \citenamefont {Andersson},
  \citenamefont {Herzog}, \citenamefont {Lane}, \citenamefont {Chantry},
  \citenamefont {Hosking},\ and\ \citenamefont {Turner}}]{vaughan2024}%
  \BibitemOpen
  \bibfield  {author} {\bibinfo {author} {\bibfnamefont {A.}~\bibnamefont
  {Vaughan}}, \bibinfo {author} {\bibfnamefont {S.}~\bibnamefont {Markou}},
  \bibinfo {author} {\bibfnamefont {W.}~\bibnamefont {Tebbutt}}, \bibinfo
  {author} {\bibfnamefont {J.}~\bibnamefont {Requeima}}, \bibinfo {author}
  {\bibfnamefont {W.~P.}\ \bibnamefont {Bruinsma}}, \bibinfo {author}
  {\bibfnamefont {T.~R.}\ \bibnamefont {Andersson}}, \bibinfo {author}
  {\bibfnamefont {M.}~\bibnamefont {Herzog}}, \bibinfo {author} {\bibfnamefont
  {N.~D.}\ \bibnamefont {Lane}}, \bibinfo {author} {\bibfnamefont
  {M.}~\bibnamefont {Chantry}}, \bibinfo {author} {\bibfnamefont {J.~S.}\
  \bibnamefont {Hosking}}, \ and\ \bibinfo {author} {\bibfnamefont {R.~E.}\
  \bibnamefont {Turner}},\ }\href {\doibase 10.48550/arXiv.2404.00411}
  {\enquote {\bibinfo {title} {Aardvark weather: end-to-end data-driven weather
  forecasting},}\ } (\bibinfo {year} {2024}),\ \Eprint
  {http://arxiv.org/abs/2404.00411} {arXiv:2404.00411 [physics.ao-ph]}
  \BibitemShut {NoStop}%
\bibitem [{\citenamefont {Bocquet}\ \emph {et~al.}(2019)\citenamefont
  {Bocquet}, \citenamefont {Brajard}, \citenamefont {Carrassi},\ and\
  \citenamefont {Bertino}}]{bocquet2019b}%
  \BibitemOpen
  \bibfield  {author} {\bibinfo {author} {\bibfnamefont {M.}~\bibnamefont
  {Bocquet}}, \bibinfo {author} {\bibfnamefont {J.}~\bibnamefont {Brajard}},
  \bibinfo {author} {\bibfnamefont {A.}~\bibnamefont {Carrassi}}, \ and\
  \bibinfo {author} {\bibfnamefont {L.}~\bibnamefont {Bertino}},\ }\bibfield
  {title} {\enquote {\bibinfo {title} {Data assimilation as a learning tool to
  infer ordinary differential equation representations of dynamical models},}\
  }\href {\doibase 10.5194/npg-26-143-2019} {\bibfield  {journal} {\bibinfo
  {journal} {Nonlin. Processes Geophys.}\ }\textbf {\bibinfo {volume} {26}},\
  \bibinfo {pages} {143--162} (\bibinfo {year} {2019})}\BibitemShut {NoStop}%
\bibitem [{\citenamefont {Brajard}\ \emph {et~al.}(2020)\citenamefont
  {Brajard}, \citenamefont {Carrassi}, \citenamefont {Bocquet},\ and\
  \citenamefont {Bertino}}]{brajard2020}%
  \BibitemOpen
  \bibfield  {author} {\bibinfo {author} {\bibfnamefont {J.}~\bibnamefont
  {Brajard}}, \bibinfo {author} {\bibfnamefont {A.}~\bibnamefont {Carrassi}},
  \bibinfo {author} {\bibfnamefont {M.}~\bibnamefont {Bocquet}}, \ and\
  \bibinfo {author} {\bibfnamefont {L.}~\bibnamefont {Bertino}},\ }\bibfield
  {title} {\enquote {\bibinfo {title} {Combining data assimilation and machine
  learning to emulate a dynamical model from sparse and noisy observations: a
  case study with the {L}orenz 96 model},}\ }\href {\doibase
  10.1016/j.jocs.2020.101171} {\bibfield  {journal} {\bibinfo  {journal} {J.
  Comput. Sci.}\ }\textbf {\bibinfo {volume} {44}},\ \bibinfo {pages} {101171}
  (\bibinfo {year} {2020})}\BibitemShut {NoStop}%
\bibitem [{\citenamefont {Bocquet}\ \emph {et~al.}(2020)\citenamefont
  {Bocquet}, \citenamefont {Brajard}, \citenamefont {Carrassi},\ and\
  \citenamefont {Bertino}}]{bocquet2020}%
  \BibitemOpen
  \bibfield  {author} {\bibinfo {author} {\bibfnamefont {M.}~\bibnamefont
  {Bocquet}}, \bibinfo {author} {\bibfnamefont {J.}~\bibnamefont {Brajard}},
  \bibinfo {author} {\bibfnamefont {A.}~\bibnamefont {Carrassi}}, \ and\
  \bibinfo {author} {\bibfnamefont {L.}~\bibnamefont {Bertino}},\ }\bibfield
  {title} {\enquote {\bibinfo {title} {Bayesian inference of chaotic dynamics
  by merging data assimilation, machine learning and
  expectation-maximization},}\ }\href {\doibase 10.3934/fods.2020004}
  {\bibfield  {journal} {\bibinfo  {journal} {Foundations of Data Science}\
  }\textbf {\bibinfo {volume} {2}},\ \bibinfo {pages} {55--80} (\bibinfo {year}
  {2020})}\BibitemShut {NoStop}%
\bibitem [{\citenamefont {Brajard}\ \emph {et~al.}(2021)\citenamefont
  {Brajard}, \citenamefont {Carrassi}, \citenamefont {Bocquet},\ and\
  \citenamefont {Bertino}}]{brajard2021}%
  \BibitemOpen
  \bibfield  {author} {\bibinfo {author} {\bibfnamefont {J.}~\bibnamefont
  {Brajard}}, \bibinfo {author} {\bibfnamefont {A.}~\bibnamefont {Carrassi}},
  \bibinfo {author} {\bibfnamefont {M.}~\bibnamefont {Bocquet}}, \ and\
  \bibinfo {author} {\bibfnamefont {L.}~\bibnamefont {Bertino}},\ }\bibfield
  {title} {\enquote {\bibinfo {title} {Combining data assimilation and machine
  learning to infer unresolved scale parametrisation},}\ }\href {\doibase
  10.1098/rsta.2020.0086} {\bibfield  {journal} {\bibinfo  {journal} {Phil.
  Trans. R. Soc. A}\ }\textbf {\bibinfo {volume} {379}},\ \bibinfo {pages}
  {20200086} (\bibinfo {year} {2021})},\ \Eprint
  {http://arxiv.org/abs/arXiv:2009.04318} {arXiv:2009.04318} \BibitemShut
  {NoStop}%
\bibitem [{\citenamefont {Liu}\ \emph {et~al.}(2022)\citenamefont {Liu},
  \citenamefont {Xu}, \citenamefont {Kurths},\ and\ \citenamefont
  {Liu}}]{liu2022}%
  \BibitemOpen
  \bibfield  {author} {\bibinfo {author} {\bibfnamefont {Q.}~\bibnamefont
  {Liu}}, \bibinfo {author} {\bibfnamefont {Y.}~\bibnamefont {Xu}}, \bibinfo
  {author} {\bibfnamefont {J.}~\bibnamefont {Kurths}}, \ and\ \bibinfo {author}
  {\bibfnamefont {X.}~\bibnamefont {Liu}},\ }\bibfield  {title} {\enquote
  {\bibinfo {title} {Complex nonlinear dynamics and vibration suppression of
  conceptual airfoil models: {A} state-of-the-art overview},}\ }\href {\doibase
  10.1063/5.0093478} {\bibfield  {journal} {\bibinfo  {journal} {Chaos}\
  }\textbf {\bibinfo {volume} {32}},\ \bibinfo {pages} {062101} (\bibinfo
  {year} {2022})}\BibitemShut {NoStop}%
\bibitem [{\citenamefont {Wang}\ \emph {et~al.}(2024)\citenamefont {Wang},
  \citenamefont {Feng}, \citenamefont {Xu},\ and\ \citenamefont
  {Kurths}}]{wang2024}%
  \BibitemOpen
  \bibfield  {author} {\bibinfo {author} {\bibfnamefont {X.}~\bibnamefont
  {Wang}}, \bibinfo {author} {\bibfnamefont {J.}~\bibnamefont {Feng}}, \bibinfo
  {author} {\bibfnamefont {Y.}~\bibnamefont {Xu}}, \ and\ \bibinfo {author}
  {\bibfnamefont {J.}~\bibnamefont {Kurths}},\ }\bibfield  {title} {\enquote
  {\bibinfo {title} {Deep learning-based state prediction of the lorenz system
  with control parameters},}\ }\href {\doibase 10.1063/5.0187866} {\bibfield
  {journal} {\bibinfo  {journal} {Chaos}\ }\textbf {\bibinfo {volume} {34}},\
  \bibinfo {pages} {033108} (\bibinfo {year} {2024})}\BibitemShut {NoStop}%
\bibitem [{\citenamefont {Misra}(2019)}]{misra2019}%
  \BibitemOpen
  \bibfield  {author} {\bibinfo {author} {\bibfnamefont {D.}~\bibnamefont
  {Misra}},\ }\bibfield  {title} {\enquote {\bibinfo {title} {Mish: {A} self
  regularized non-monotonic neural activation function},}\ }\href {\doibase
  10.48550/arXiv.1908.08681} {\bibfield  {journal} {\bibinfo  {journal}
  {arXiv}\ } (\bibinfo {year} {2019}),\ 10.48550/arXiv.1908.08681}\BibitemShut
  {NoStop}%
\bibitem [{\citenamefont {Kingma}\ and\ \citenamefont {Ba}(2015)}]{kingma2015}%
  \BibitemOpen
  \bibfield  {author} {\bibinfo {author} {\bibfnamefont {D.~P.}\ \bibnamefont
  {Kingma}}\ and\ \bibinfo {author} {\bibfnamefont {J.}~\bibnamefont {Ba}},\
  }\bibfield  {title} {\enquote {\bibinfo {title} {Adam: {A} method for
  stochastic optimization},}\ }in\ \href {\doibase 10.48550/arXiv.1412.6980}
  {\emph {\bibinfo {booktitle} {International Conference on Learning
  Representations (ICLR)}}}\ (\bibinfo {address} {San Diega, CA, USA},\
  \bibinfo {year} {2015})\BibitemShut {NoStop}%
\bibitem [{\citenamefont {Tang}\ and\ \citenamefont {Glass}(2018)}]{tang2018}%
  \BibitemOpen
  \bibfield  {author} {\bibinfo {author} {\bibfnamefont {H.}~\bibnamefont
  {Tang}}\ and\ \bibinfo {author} {\bibfnamefont {J.}~\bibnamefont {Glass}},\
  }\bibfield  {title} {\enquote {\bibinfo {title} {On training recurrent
  networks with truncated backpropagation through time in speech
  recognition},}\ }in\ \href {\doibase 10.1109/SLT.2018.8639517} {\emph
  {\bibinfo {booktitle} {2018 IEEE Spoken Language Technology Workshop
  (SLT)}}}\ (\bibinfo {year} {2018})\ pp.\ \bibinfo {pages}
  {48--55}\BibitemShut {NoStop}%
\bibitem [{\citenamefont {Aicher}, \citenamefont {Foti},\ and\ \citenamefont
  {Fox}(2020)}]{aicher2020}%
  \BibitemOpen
  \bibfield  {author} {\bibinfo {author} {\bibfnamefont {C.}~\bibnamefont
  {Aicher}}, \bibinfo {author} {\bibfnamefont {N.~J.}\ \bibnamefont {Foti}}, \
  and\ \bibinfo {author} {\bibfnamefont {E.~B.}\ \bibnamefont {Fox}},\
  }\bibfield  {title} {\enquote {\bibinfo {title} {Adaptively truncating
  backpropagation through time to control gradient bias},}\ }in\ \href
  {https://proceedings.mlr.press/v115/aicher20a.html} {\emph {\bibinfo
  {booktitle} {Proceedings of The 35th Uncertainty in Artificial Intelligence
  Conference}}},\ \bibinfo {series} {Proceedings of Machine Learning Research},
  Vol.\ \bibinfo {volume} {115},\ \bibinfo {editor} {edited by\ \bibinfo
  {editor} {\bibfnamefont {R.~P.}\ \bibnamefont {Adams}}\ and\ \bibinfo
  {editor} {\bibfnamefont {V.}~\bibnamefont {Gogate}}}\ (\bibinfo  {publisher}
  {PMLR},\ \bibinfo {year} {2020})\ pp.\ \bibinfo {pages}
  {799--808}\BibitemShut {NoStop}%
\bibitem [{\citenamefont {Carrassi}\ \emph
  {et~al.}(2008{\natexlab{b}})\citenamefont {Carrassi}, \citenamefont {Ghil},
  \citenamefont {Trevisan},\ and\ \citenamefont {Uboldi}}]{carrassi2008a}%
  \BibitemOpen
  \bibfield  {author} {\bibinfo {author} {\bibfnamefont {A.}~\bibnamefont
  {Carrassi}}, \bibinfo {author} {\bibfnamefont {M.}~\bibnamefont {Ghil}},
  \bibinfo {author} {\bibfnamefont {A.}~\bibnamefont {Trevisan}}, \ and\
  \bibinfo {author} {\bibfnamefont {F.}~\bibnamefont {Uboldi}},\ }\bibfield
  {title} {\enquote {\bibinfo {title} {Data assimilation as a nonlinear
  dynamical systems problem: Stability and convergence of the
  prediction-assimilation system},}\ }\href {\doibase 10.1063/1.2909862}
  {\bibfield  {journal} {\bibinfo  {journal} {Chaos}\ }\textbf {\bibinfo
  {volume} {18}},\ \bibinfo {pages} {023112} (\bibinfo {year}
  {2008}{\natexlab{b}})}\BibitemShut {NoStop}%
\bibitem [{\citenamefont {Carrassi}\ \emph
  {et~al.}(2008{\natexlab{c}})\citenamefont {Carrassi}, \citenamefont
  {Trevisan}, \citenamefont {Descamps}, \citenamefont {Talagrand},\ and\
  \citenamefont {Uboldi}}]{carrassi2008b}%
  \BibitemOpen
  \bibfield  {author} {\bibinfo {author} {\bibfnamefont {A.}~\bibnamefont
  {Carrassi}}, \bibinfo {author} {\bibfnamefont {A.}~\bibnamefont {Trevisan}},
  \bibinfo {author} {\bibfnamefont {L.}~\bibnamefont {Descamps}}, \bibinfo
  {author} {\bibfnamefont {O.}~\bibnamefont {Talagrand}}, \ and\ \bibinfo
  {author} {\bibfnamefont {F.}~\bibnamefont {Uboldi}},\ }\bibfield  {title}
  {\enquote {\bibinfo {title} {Controlling instabilities along a {3DVar}
  analysis cycle by assimilating in the unstable subspace: a comparison with
  the {EnKF}},}\ }\href {\doibase 10.5194/npg-15-503-2008} {\bibfield
  {journal} {\bibinfo  {journal} {Nonlin. Processes Geophys.}\ }\textbf
  {\bibinfo {volume} {15}},\ \bibinfo {pages} {503--521} (\bibinfo {year}
  {2008}{\natexlab{c}})}\BibitemShut {NoStop}%
\bibitem [{\citenamefont {Goodfellow}, \citenamefont {Bengio},\ and\
  \citenamefont {Courville}(2016)}]{goodfellow2016}%
  \BibitemOpen
  \bibfield  {author} {\bibinfo {author} {\bibfnamefont {I.}~\bibnamefont
  {Goodfellow}}, \bibinfo {author} {\bibfnamefont {Y.}~\bibnamefont {Bengio}},
  \ and\ \bibinfo {author} {\bibfnamefont {A.}~\bibnamefont {Courville}},\
  }\href@noop {} {\emph {\bibinfo {title} {Deep learning}}}\ (\bibinfo
  {publisher} {The MIT Press, Cambridge Massachusetts, London England},\
  \bibinfo {year} {2016})\ p.\ \bibinfo {pages} {775}\BibitemShut {NoStop}%
\bibitem [{\citenamefont {Bocquet}, \citenamefont {Raanes},\ and\ \citenamefont
  {Hannart}(2015)}]{bocquet2015b}%
  \BibitemOpen
  \bibfield  {author} {\bibinfo {author} {\bibfnamefont {M.}~\bibnamefont
  {Bocquet}}, \bibinfo {author} {\bibfnamefont {P.~N.}\ \bibnamefont {Raanes}},
  \ and\ \bibinfo {author} {\bibfnamefont {A.}~\bibnamefont {Hannart}},\
  }\bibfield  {title} {\enquote {\bibinfo {title} {Expanding the validity of
  the ensemble {K}alman filter without the intrinsic need for inflation},}\
  }\href {\doibase 10.5194/npg-22-645-2015} {\bibfield  {journal} {\bibinfo
  {journal} {Nonlin. Processes Geophys.}\ }\textbf {\bibinfo {volume} {22}},\
  \bibinfo {pages} {645--662} (\bibinfo {year} {2015})}\BibitemShut {NoStop}%
\bibitem [{\citenamefont {Raanes}, \citenamefont {Bocquet},\ and\ \citenamefont
  {Carrassi}(2019)}]{raanes2019a}%
  \BibitemOpen
  \bibfield  {author} {\bibinfo {author} {\bibfnamefont {P.~N.}\ \bibnamefont
  {Raanes}}, \bibinfo {author} {\bibfnamefont {M.}~\bibnamefont {Bocquet}}, \
  and\ \bibinfo {author} {\bibfnamefont {A.}~\bibnamefont {Carrassi}},\
  }\bibfield  {title} {\enquote {\bibinfo {title} {Adaptive covariance
  inflation in the ensemble {K}alman filter by {G}aussian scale mixtures},}\
  }\href {\doibase 10.1002/qj.3386} {\bibfield  {journal} {\bibinfo  {journal}
  {Q. J. R. Meteorol. Soc.}\ }\textbf {\bibinfo {volume} {145}},\ \bibinfo
  {pages} {53--75} (\bibinfo {year} {2019})}\BibitemShut {NoStop}%
\bibitem [{\citenamefont {van Amersfoort}\ \emph {et~al.}(2022)\citenamefont
  {van Amersfoort}, \citenamefont {Smith}, \citenamefont {Jesson},
  \citenamefont {Key},\ and\ \citenamefont {Gal}}]{vanamersfoort2022}%
  \BibitemOpen
  \bibfield  {author} {\bibinfo {author} {\bibfnamefont {J.}~\bibnamefont {van
  Amersfoort}}, \bibinfo {author} {\bibfnamefont {L.}~\bibnamefont {Smith}},
  \bibinfo {author} {\bibfnamefont {A.}~\bibnamefont {Jesson}}, \bibinfo
  {author} {\bibfnamefont {O.}~\bibnamefont {Key}}, \ and\ \bibinfo {author}
  {\bibfnamefont {Y.}~\bibnamefont {Gal}},\ }\href {\doibase
  10.48550/arXiv.2102.11409} {\enquote {\bibinfo {title} {On feature collapse
  and deep kernel learning for single forward pass uncertainty},}\ } (\bibinfo
  {year} {2022}),\ \Eprint {http://arxiv.org/abs/2102.11409} {arXiv:2102.11409
  [cs.LG]} \BibitemShut {NoStop}%
\bibitem [{\citenamefont {Peyron}\ \emph {et~al.}(2021)\citenamefont {Peyron},
  \citenamefont {Fillion}, \citenamefont {G{\"u}rol}, \citenamefont {Marchais},
  \citenamefont {Gratton}, \citenamefont {Boudier},\ and\ \citenamefont
  {Goret}}]{peyron2021}%
  \BibitemOpen
  \bibfield  {author} {\bibinfo {author} {\bibfnamefont {M.}~\bibnamefont
  {Peyron}}, \bibinfo {author} {\bibfnamefont {A.}~\bibnamefont {Fillion}},
  \bibinfo {author} {\bibfnamefont {S.}~\bibnamefont {G{\"u}rol}}, \bibinfo
  {author} {\bibfnamefont {V.}~\bibnamefont {Marchais}}, \bibinfo {author}
  {\bibfnamefont {S.}~\bibnamefont {Gratton}}, \bibinfo {author} {\bibfnamefont
  {P.}~\bibnamefont {Boudier}}, \ and\ \bibinfo {author} {\bibfnamefont
  {G.}~\bibnamefont {Goret}},\ }\bibfield  {title} {\enquote {\bibinfo {title}
  {Latent space data assimilation by using deep learning},}\ }\href {\doibase
  10.1002/qj.4153} {\bibfield  {journal} {\bibinfo  {journal} {Q. J. R.
  Meteorol. Soc.}\ }\textbf {\bibinfo {volume} {147}},\ \bibinfo {pages}
  {3759--3777} (\bibinfo {year} {2021})}\BibitemShut {NoStop}%
\bibitem [{\citenamefont {Kalman}(1960)}]{kalman1960}%
  \BibitemOpen
  \bibfield  {author} {\bibinfo {author} {\bibfnamefont {R.~E.}\ \bibnamefont
  {Kalman}},\ }\bibfield  {title} {\enquote {\bibinfo {title} {A new approach
  to linear filtering and prediction problems},}\ }\href {\doibase
  10.1115/1.3662552} {\bibfield  {journal} {\bibinfo  {journal} {J. Basic
  Eng.}\ }\textbf {\bibinfo {volume} {82}},\ \bibinfo {pages} {35--45}
  (\bibinfo {year} {1960})}\BibitemShut {NoStop}%
\bibitem [{\citenamefont {Ghil}\ and\ \citenamefont
  {Malanotte-Rizzoli}(1991)}]{ghil1991}%
  \BibitemOpen
  \bibfield  {author} {\bibinfo {author} {\bibfnamefont {M.}~\bibnamefont
  {Ghil}}\ and\ \bibinfo {author} {\bibfnamefont {P.}~\bibnamefont
  {Malanotte-Rizzoli}},\ }\bibfield  {title} {\enquote {\bibinfo {title} {Data
  assimilation in meteorology and oceanography},}\ }\href {\doibase
  10.1016/S0065-2687(08)60442-2} {\bibfield  {journal} {\bibinfo  {journal}
  {Advanc. in Geophys.}\ }\textbf {\bibinfo {volume} {33}},\ \bibinfo {pages}
  {141--266} (\bibinfo {year} {1991})}\BibitemShut {NoStop}%
\bibitem [{\citenamefont {Bhatia}, \citenamefont {Jain},\ and\ \citenamefont
  {Lim}(2019)}]{bhatia2019a}%
  \BibitemOpen
  \bibfield  {author} {\bibinfo {author} {\bibfnamefont {R.}~\bibnamefont
  {Bhatia}}, \bibinfo {author} {\bibfnamefont {T.}~\bibnamefont {Jain}}, \ and\
  \bibinfo {author} {\bibfnamefont {Y.}~\bibnamefont {Lim}},\ }\bibfield
  {title} {\enquote {\bibinfo {title} {On the {B}ures-{W}asserstein distance
  between positive definite matrices},}\ }\href {\doibase
  10.1016/j.exmath.2018.01.002} {\bibfield  {journal} {\bibinfo  {journal}
  {Expo. Math.}\ }\textbf {\bibinfo {volume} {37}},\ \bibinfo {pages}
  {165--191} (\bibinfo {year} {2019})}\BibitemShut {NoStop}%
\bibitem [{\citenamefont {Oseledec}(1968)}]{oseledec1968}%
  \BibitemOpen
  \bibfield  {author} {\bibinfo {author} {\bibfnamefont {V.~I.}\ \bibnamefont
  {Oseledec}},\ }\bibfield  {title} {\enquote {\bibinfo {title} {A
  multiplicative ergodic theorem. ljapunov characteristic numbers for dynamical
  systems.}}\ }\href@noop {} {\bibfield  {journal} {\bibinfo  {journal} {Trans.
  Moscow Math Soc.}\ }\textbf {\bibinfo {volume} {19}},\ \bibinfo {pages}
  {197--231} (\bibinfo {year} {1968})}\BibitemShut {NoStop}%
\bibitem [{\citenamefont {Arnold}(1998)}]{arnold1998}%
  \BibitemOpen
  \bibfield  {author} {\bibinfo {author} {\bibfnamefont {L.}~\bibnamefont
  {Arnold}},\ }\href {\doibase 10.1007/978-3-662-12878-7} {\emph {\bibinfo
  {title} {Random Dynamical Systems}}}\ (\bibinfo  {publisher} {Springer
  Berlin, Heidelberg},\ \bibinfo {year} {1998})\ p.\ \bibinfo {pages}
  {586}\BibitemShut {NoStop}%
\bibitem [{\citenamefont {Chekroun}, \citenamefont {Simonnet},\ and\
  \citenamefont {Ghil}(2011)}]{chekroun2011}%
  \BibitemOpen
  \bibfield  {author} {\bibinfo {author} {\bibfnamefont {M.~D.}\ \bibnamefont
  {Chekroun}}, \bibinfo {author} {\bibfnamefont {E.}~\bibnamefont {Simonnet}},
  \ and\ \bibinfo {author} {\bibfnamefont {M.}~\bibnamefont {Ghil}},\
  }\bibfield  {title} {\enquote {\bibinfo {title} {Stochastic climate dynamics:
  {R}andom attractors and time-dependent invariant measures},}\ }\href
  {\doibase 10.1016/j.physd.2011.06.005} {\bibfield  {journal} {\bibinfo
  {journal} {Physica D}\ }\textbf {\bibinfo {volume} {240}},\ \bibinfo {pages}
  {1685--1700} (\bibinfo {year} {2011})}\BibitemShut {NoStop}%
\bibitem [{\citenamefont {Flandoli}\ and\ \citenamefont
  {Tonello}(2021)}]{flandoli2021}%
  \BibitemOpen
  \bibfield  {author} {\bibinfo {author} {\bibfnamefont {F.}~\bibnamefont
  {Flandoli}}\ and\ \bibinfo {author} {\bibfnamefont {E.}~\bibnamefont
  {Tonello}},\ }\href {https://pagine.dm.unipi.it/flandoli/Part1bis.pdf}
  {\enquote {\bibinfo {title} {An introduction to random dynamical systems for
  climate},}\ } (\bibinfo {year} {2021})\BibitemShut {NoStop}%
\bibitem [{\citenamefont {Ghil}\ and\ \citenamefont
  {Sciamarella}(2023)}]{ghil2023}%
  \BibitemOpen
  \bibfield  {author} {\bibinfo {author} {\bibfnamefont {M.}~\bibnamefont
  {Ghil}}\ and\ \bibinfo {author} {\bibfnamefont {D.}~\bibnamefont
  {Sciamarella}},\ }\bibfield  {title} {\enquote {\bibinfo {title} {Review
  article: {D}ynamical systems, algebraic topology and the climate sciences},}\
  }\href {\doibase 10.5194/npg-30-399-2023} {\bibfield  {journal} {\bibinfo
  {journal} {Nonlin. Processes Geophys.}\ ,\ \bibinfo {pages} {399--434}}
  (\bibinfo {year} {2023})}\BibitemShut {NoStop}%
\bibitem [{\citenamefont {Paz{\'o}}, \citenamefont {Rodr{\'i}guez},\ and\
  \citenamefont {L{\'o}pez}(2010)}]{pazo2010}%
  \BibitemOpen
  \bibfield  {author} {\bibinfo {author} {\bibfnamefont {D.}~\bibnamefont
  {Paz{\'o}}}, \bibinfo {author} {\bibfnamefont {M.~A.}\ \bibnamefont
  {Rodr{\'i}guez}}, \ and\ \bibinfo {author} {\bibfnamefont {J.~M.}\
  \bibnamefont {L{\'o}pez}},\ }\bibfield  {title} {\enquote {\bibinfo {title}
  {Spatio-temporal evolution of perturbations in ensembles initialized by bred,
  {L}yapunov and singular vectors},}\ }\href {\doibase
  10.1111/j.1600-0870.2009.00419.x} {\bibfield  {journal} {\bibinfo  {journal}
  {Tellus A}\ }\textbf {\bibinfo {volume} {62}},\ \bibinfo {pages} {10--23}
  (\bibinfo {year} {2010})}\BibitemShut {NoStop}%
\bibitem [{\citenamefont {Vannitsem}\ and\ \citenamefont
  {Lucarini}(2016)}]{vannitsem2016}%
  \BibitemOpen
  \bibfield  {author} {\bibinfo {author} {\bibfnamefont {S.}~\bibnamefont
  {Vannitsem}}\ and\ \bibinfo {author} {\bibfnamefont {V.}~\bibnamefont
  {Lucarini}},\ }\bibfield  {title} {\enquote {\bibinfo {title} {Statistical
  and dynamical properties of covariant {L}yapunov vectors in a coupled
  atmosphere-ocean model-multiscale effects, geometric degeneracy, and error
  dynamics},}\ }\href {\doibase 10.1088/1751-8113/49/22/224001} {\bibfield
  {journal} {\bibinfo  {journal} {J. Phys. A: Math. Theor.}\ }\textbf {\bibinfo
  {volume} {49}},\ \bibinfo {pages} {224001} (\bibinfo {year}
  {2016})}\BibitemShut {NoStop}%
\bibitem [{\citenamefont {Giggins}\ and\ \citenamefont
  {Gottwald}(2019)}]{giggins2019}%
  \BibitemOpen
  \bibfield  {author} {\bibinfo {author} {\bibfnamefont {B.}~\bibnamefont
  {Giggins}}\ and\ \bibinfo {author} {\bibfnamefont {G.~A.}\ \bibnamefont
  {Gottwald}},\ }\bibfield  {title} {\enquote {\bibinfo {title} {Stochastically
  perturbed bred vectors in multi-scale systems},}\ }\href {\doibase
  10.1002/qj.3457} {\bibfield  {journal} {\bibinfo  {journal} {Q. J. R.
  Meteorol. Soc.}\ }\textbf {\bibinfo {volume} {145}},\ \bibinfo {pages}
  {642--658} (\bibinfo {year} {2019})}\BibitemShut {NoStop}%
\bibitem [{\citenamefont {Kuramoto}\ and\ \citenamefont
  {Tsuzuki}(1976)}]{kuramoto1976}%
  \BibitemOpen
  \bibfield  {author} {\bibinfo {author} {\bibfnamefont {Y.}~\bibnamefont
  {Kuramoto}}\ and\ \bibinfo {author} {\bibfnamefont {T.}~\bibnamefont
  {Tsuzuki}},\ }\bibfield  {title} {\enquote {\bibinfo {title} {Persistent
  propagation of concentration waves in dissipative media far from thermal
  equilibrium},}\ }\href {\doibase 10.1143/PTP.55.356} {\bibfield  {journal}
  {\bibinfo  {journal} {Progr. Theoret. Phys.}\ }\textbf {\bibinfo {volume}
  {55}},\ \bibinfo {pages} {356--369} (\bibinfo {year} {1976})}\BibitemShut
  {NoStop}%
\bibitem [{\citenamefont {Sivashinsky}(1977)}]{sivashinsky1977}%
  \BibitemOpen
  \bibfield  {author} {\bibinfo {author} {\bibfnamefont {G.~I.}\ \bibnamefont
  {Sivashinsky}},\ }\bibfield  {title} {\enquote {\bibinfo {title} {Nonlinear
  analysis of hydrodynamic instability in laminar flames-{I}. {D}erivation of
  basic equations},}\ }\href {\doibase 10.1016/0094-5765(77)90096-0} {\bibfield
   {journal} {\bibinfo  {journal} {Acta Astronaut.}\ }\textbf {\bibinfo
  {volume} {4}},\ \bibinfo {pages} {1177--1206} (\bibinfo {year}
  {1977})}\BibitemShut {NoStop}%
\bibitem [{\citenamefont {Pannekoucke}\ \emph {et~al.}(2016)\citenamefont
  {Pannekoucke}, \citenamefont {Ricci}, \citenamefont {Barthelemy},
  \citenamefont {M{\'e}nard},\ and\ \citenamefont {Thual}}]{pannekoucke2016}%
  \BibitemOpen
  \bibfield  {author} {\bibinfo {author} {\bibfnamefont {O.}~\bibnamefont
  {Pannekoucke}}, \bibinfo {author} {\bibfnamefont {S.}~\bibnamefont {Ricci}},
  \bibinfo {author} {\bibfnamefont {S.}~\bibnamefont {Barthelemy}}, \bibinfo
  {author} {\bibfnamefont {R.}~\bibnamefont {M{\'e}nard}}, \ and\ \bibinfo
  {author} {\bibfnamefont {O.}~\bibnamefont {Thual}},\ }\bibfield  {title}
  {\enquote {\bibinfo {title} {Parametric {K}alman filter for chemical
  transport model},}\ }\href {\doibase 10.3402/tellusa.v68.31547} {\bibfield
  {journal} {\bibinfo  {journal} {Tellus A}\ }\textbf {\bibinfo {volume}
  {68}},\ \bibinfo {pages} {31457} (\bibinfo {year} {2016})}\BibitemShut
  {NoStop}%
\bibitem [{\citenamefont {Pannekoucke}, \citenamefont {Bocquet},\ and\
  \citenamefont {M{\'e}nard}(2018)}]{pannekoucke2018}%
  \BibitemOpen
  \bibfield  {author} {\bibinfo {author} {\bibfnamefont {O.}~\bibnamefont
  {Pannekoucke}}, \bibinfo {author} {\bibfnamefont {M.}~\bibnamefont
  {Bocquet}}, \ and\ \bibinfo {author} {\bibfnamefont {R.}~\bibnamefont
  {M{\'e}nard}},\ }\bibfield  {title} {\enquote {\bibinfo {title} {Parametric
  covariance dynamics for the nonlinear diffusive {B}urgers equation},}\ }\href
  {\doibase 10.5194/npg-25-481-2018} {\bibfield  {journal} {\bibinfo  {journal}
  {Nonlin. Processes Geophys.}\ }\textbf {\bibinfo {volume} {25}},\ \bibinfo
  {pages} {481--495} (\bibinfo {year} {2018})}\BibitemShut {NoStop}%
\end{thebibliography}%

\newpage

\onecolumngrid
\section*{Supplementary Material: Accurate deep learning-based filtering for chaotic dynamics by identifying
  instabilities without an ensemble}
\twocolumngrid

In the following DA, DL, CNN, EnKF, RMSE and aRMSE are the initialisms and acronyms of data assimilation, deep learning,
convolutional neural network, ensemble Kalman filter, root mean square error (of the analysis state versus the true
state), and time-averaged root mean square error (of the analysis state versus the true state), respectively. All the
$\dan$ operators trained and tested here are based on $\Ne=1$, i.e. no ensemble but a single state forecast.

\subsection{Impact of the training dataset size on the DA tests}
\label{sec:training_dataset_size}

\begin{figure}[ht]
  \includegraphics[width=0.45\textwidth]{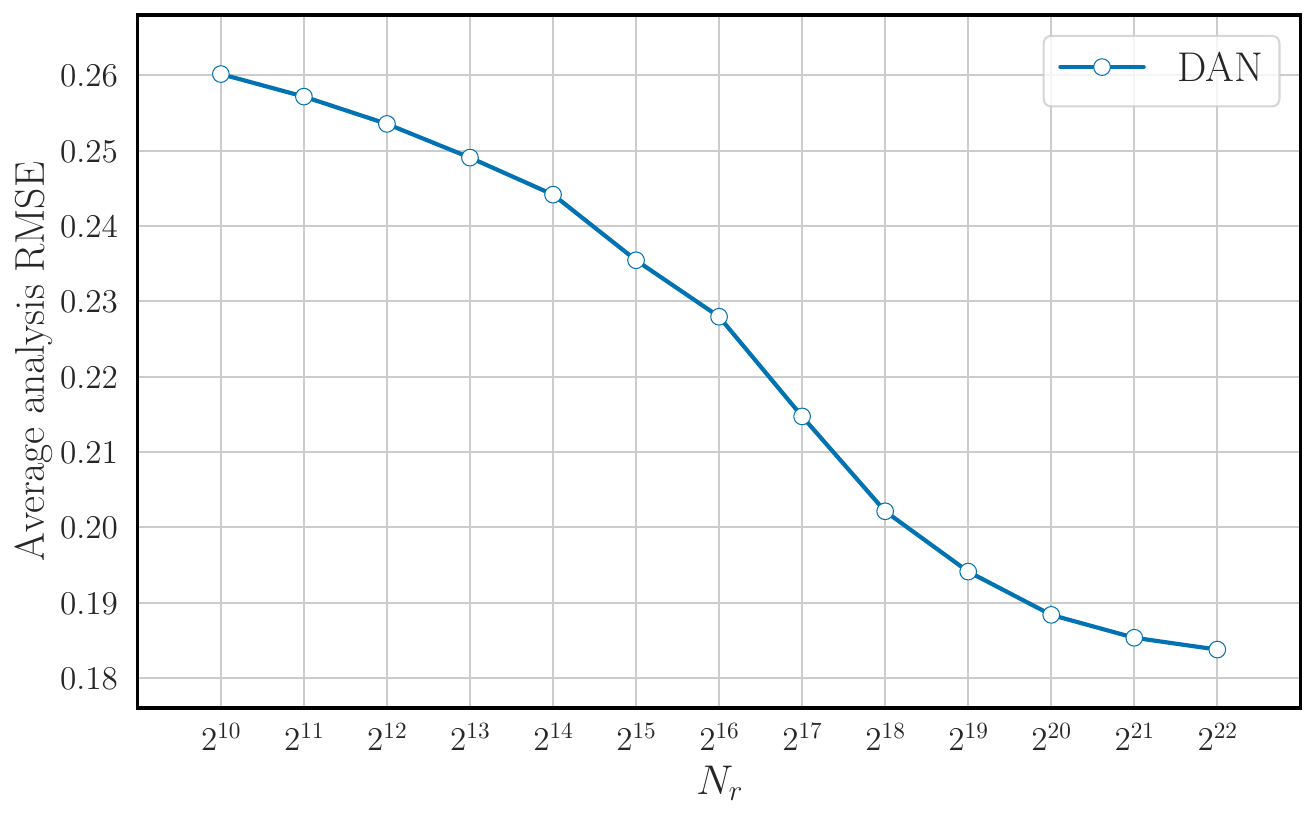}
  \caption{ \label{fig:dan_perf_Nr_l96} Test aRMSEs of $\dan$ operators learned from a L96 models with a varying
    number of trajectories $\Nr$ in the training dataset, commensurate to the size of the training dataset per epoch.}
\end{figure}

A series of $\dan$ operators are learned from increasingly larger training and validation datasets (per epoch), which is achieved by
increasing the number of trajectories $\Nr$ in the datasets in the range $[2^{10}, 2^{22}]$, and displayed in
Fig.~\ref{fig:dan_perf_Nr_l96}. The hyperparameters of the CNN architecture are those of the reference configuration.
The batch size, i.e. the number of trajectories in the training batch, is chosen in this experiment to be
$\mathrm{min}\(\Nr/10,2^{11}=2,048\)$ in order to be able to maintain the targeted training and validation datasets ratio,
while seeking a large batch for numerical efficiency.  As a compromise between accuracy and computational time for this
study, most of the experiments are conducted with $\Nr=2^{18}=262,144$.

\subsection{Impact of the number of filters of $\dan$ as a CNN on the DA tests}
\label{sec:number_of_channels}

A series of $\dan$ operators are learned from an increasingly larger number of channels/filters $\Nf$. All other hyperparameters
of the CNN architecture are those of the reference configuration.
\begin{figure}[ht]
  \includegraphics[width=0.45\textwidth]{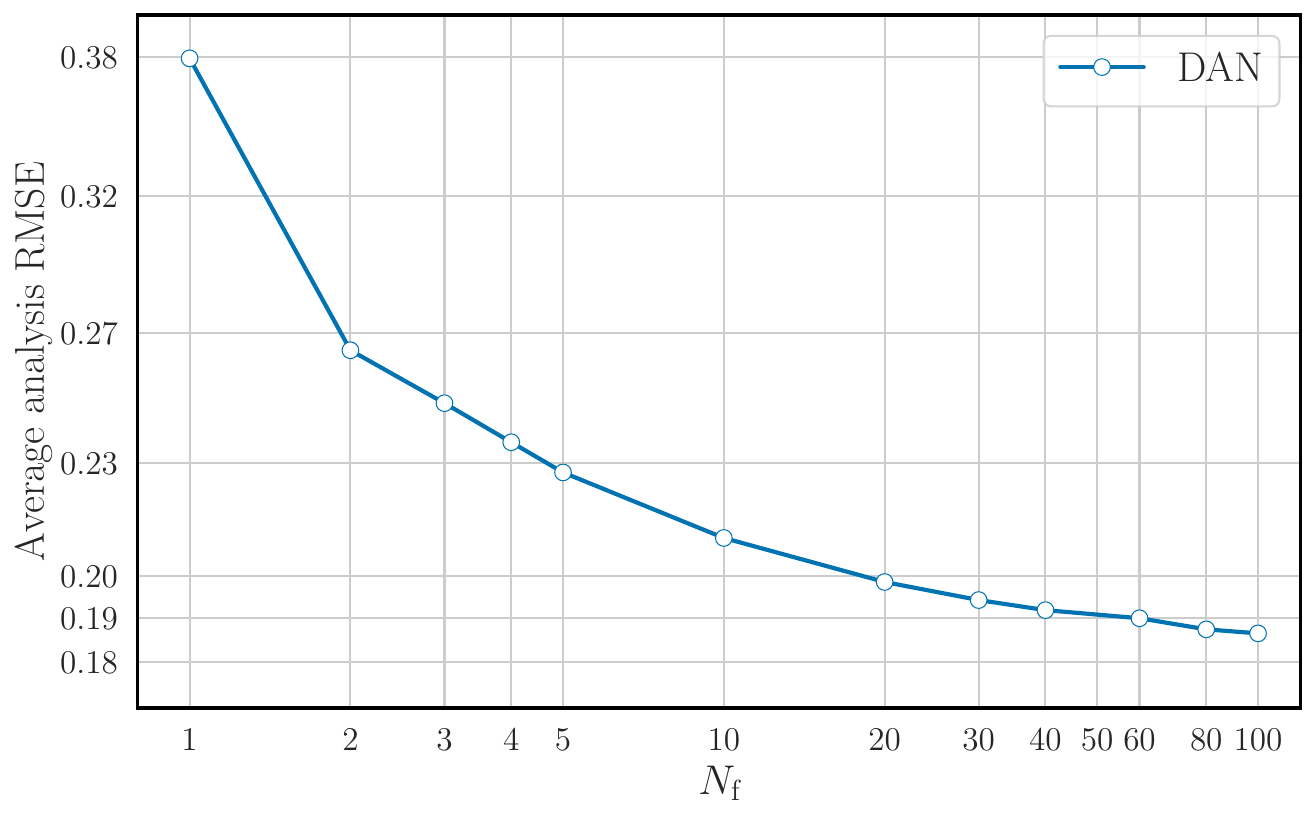}
  \caption{ \label{fig:dan_perf_Nf_l96} Test aRMSEs of the DAN method applied to the L96 model, as a function of the
    number of channels/filters $\Nf$ in $\dan$.}
\end{figure}
As a compromise between accuracy and computational time for this paper, most of the experiments are conducted with
$\Nf=40$.

\subsection{Impact of the observation noise magnitude on the DA tests}
\label{sec:observation_noise}

\begin{figure}[ht]
  \includegraphics[width=0.45\textwidth]{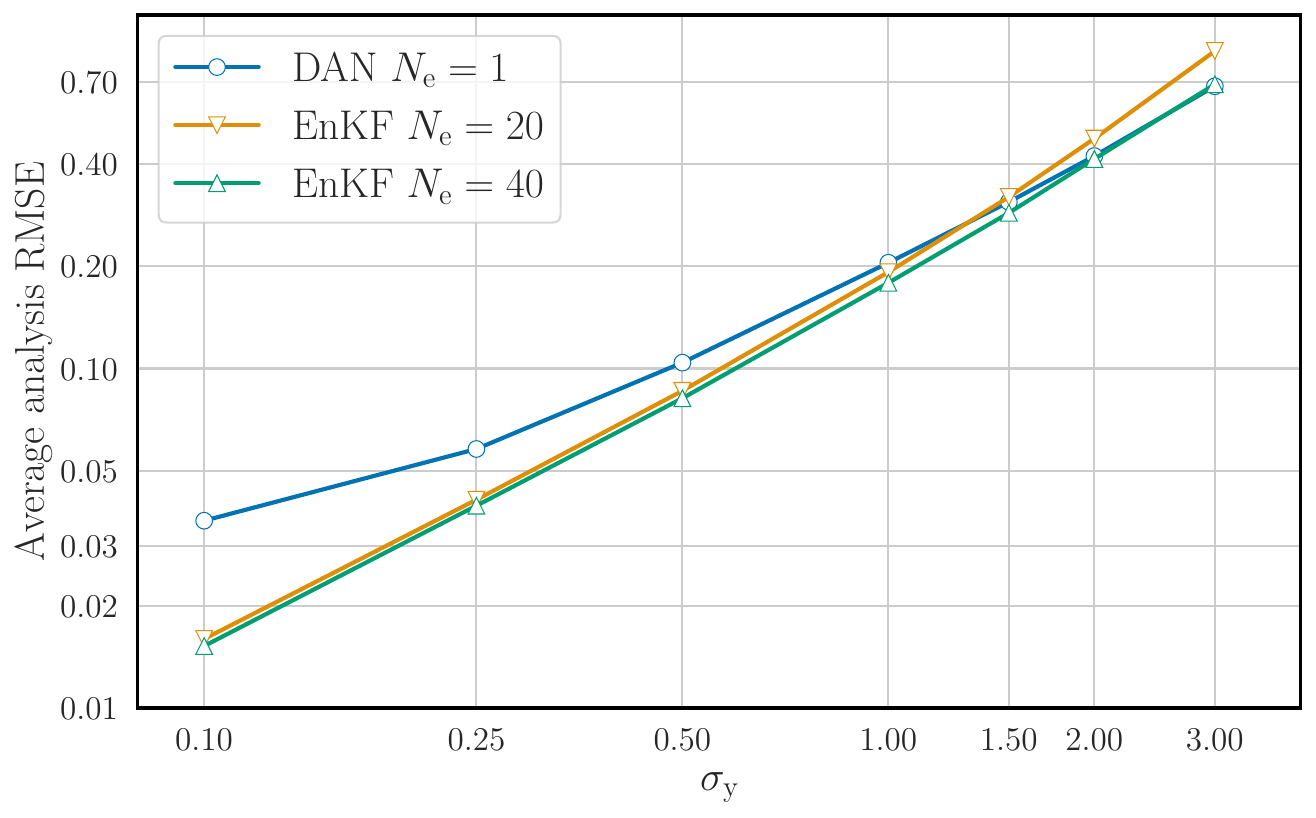}
  \caption{ \label{fig:dan_perf_sigy_l96} Test aRMSEs of the DAN method, and of the EnKF $\Ne=20,40$ applied to the L96
    model, as a function of the true standard deviation of the assimilated observations.}
\end{figure}

The impact of observation noise in the DA test experiments with magnitude distinct from the noise magnitude used in the
training of $\dan$ is discussed in the paper in Section III.B.  Figure~\ref{fig:dan_perf_sigy_l96} displays the
corresponding DAN and EnKF aRMSE curves. In all cases the analysis aRMSE remains below the corresponding observational
error standard deviation $\sigy$.

\subsection{Impact of the sparsity of the observations on the DA tests}
\label{sec:observation_sparsity}

\begin{figure}[ht]
  \includegraphics[width=0.45\textwidth]{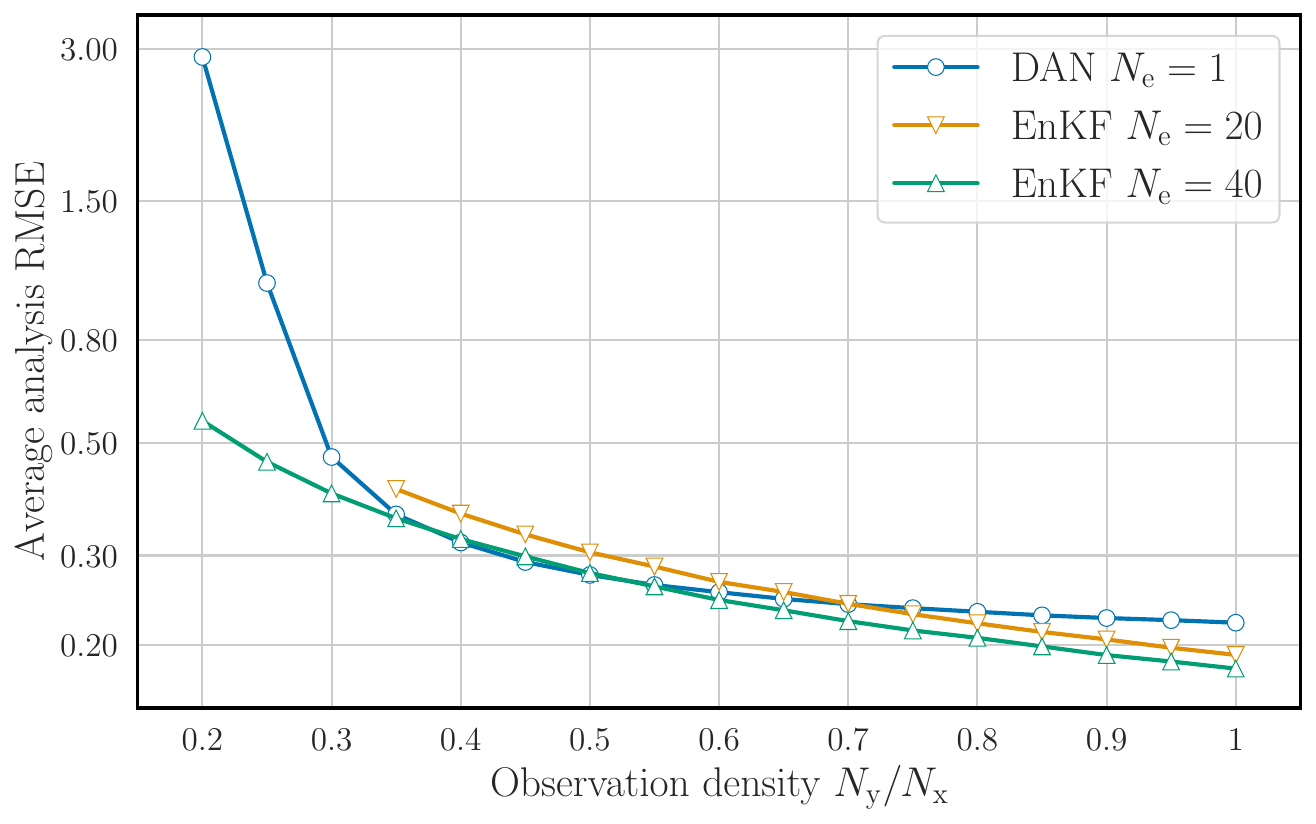}
  \caption{ \label{fig:dan_perf_obs_density_l96} Test aRMSEs of the DAN method, and of the EnKF $\Ne=20,40$ applied to
    the L96 model, as a function of the observation density $\Ny/\Nx$.}
\end{figure}

The impact on the test aRMSEs of the DAN scheme with non-trivial observation operators $\Hc_k$ is discussed in the paper
in Section III.B, with an experiment where $\dan$ is trained with varying $\Hc_k$ of time-dependent random sparsity
ratios, and then tested with observation operators of fixed in time sparsity
ratio. Figure~\ref{fig:dan_perf_obs_density_l96} displays the corresponding DAN and EnKF aRMSE curves. The observation
error standard deviation is $\sigy=1$.
Moreover, we report here (though not in the paper) that, as a sanity check, we learned $\dan$ and tested the resulting
DAN scheme with a staggered observation operator, where only the variables of the odd sites of the L96 model are
observed (a classical test in methodological DA studies). We obtain for the DL-based DA scheme and a well-tuned EnKF
with $\Ne=40$, aRMSEs of $0.289$ and $0.288$, respectively, which suggests that non-trivial observation operators are
properly dealt with by $\dan$.

\subsection{Key aRMSE scores of the main DA methods}

For comparison and for reference, the aRMSE scores of the classical DA methods and of the DL-based DAN schemes tested in
this paper, are compiled in Tab.~\ref{tab:scores}. The tested well-tuned EnKFs are based on the finite-size EnKF
(EnKF-N) DA method with has implicit adaptive inflation and relies here on an ensemble size of either $\Ne=20$ or
$\Ne=40$, beyond the unstable-neutral subspace dimension $\Nu=14$ where localization is unnecessary. The architecture
and hyperparameters of the DL-based DA schemes are those of the reference configuration described in the paper, with
possibly the exception of $\Ne$ and $\Nf$ which are hence explicitly specified in the table.

\begin{table}[ht]
\caption{\label{tab:scores} Compilation of the key aRMSE scores of the classical and DL-based DA methods tested in this paper.}
\begin{tabular}{lccc}
\hline\hline
  DA method & well-tuned classical & DL-based & aRMSE \\
  \hline
  EnKF-N, $\Ne=20$ & yes & & $0.191$ \\
  EnKF-N, $\Ne=40$ & yes & &  $0.179$ \\
  3D-Var & yes & & $0.40$ \\
  \hline
  $\dan$, $\Ne=1$, $\Nf=40$ &  & yes & $0.191$ \\
  $\dan$, $\Ne=1$, $\Nf=100$ & & yes & $0.185$ \\
  linear $\dan$, $\Ne=1$, $\Nf=40$ & & yes & $0.384$ \\
  simplified $\hdan$, $\Ne=1$, $\Nf=40$ & & yes & $0.382$
  \\
  \hline\hline
\end{tabular}
\end{table}

\subsection{Typical RMSE time series of DAN and of a well-tuned EnKF}
\label{sec:time_series}

\begin{figure}[ht]
  \includegraphics[width=0.45\textwidth]{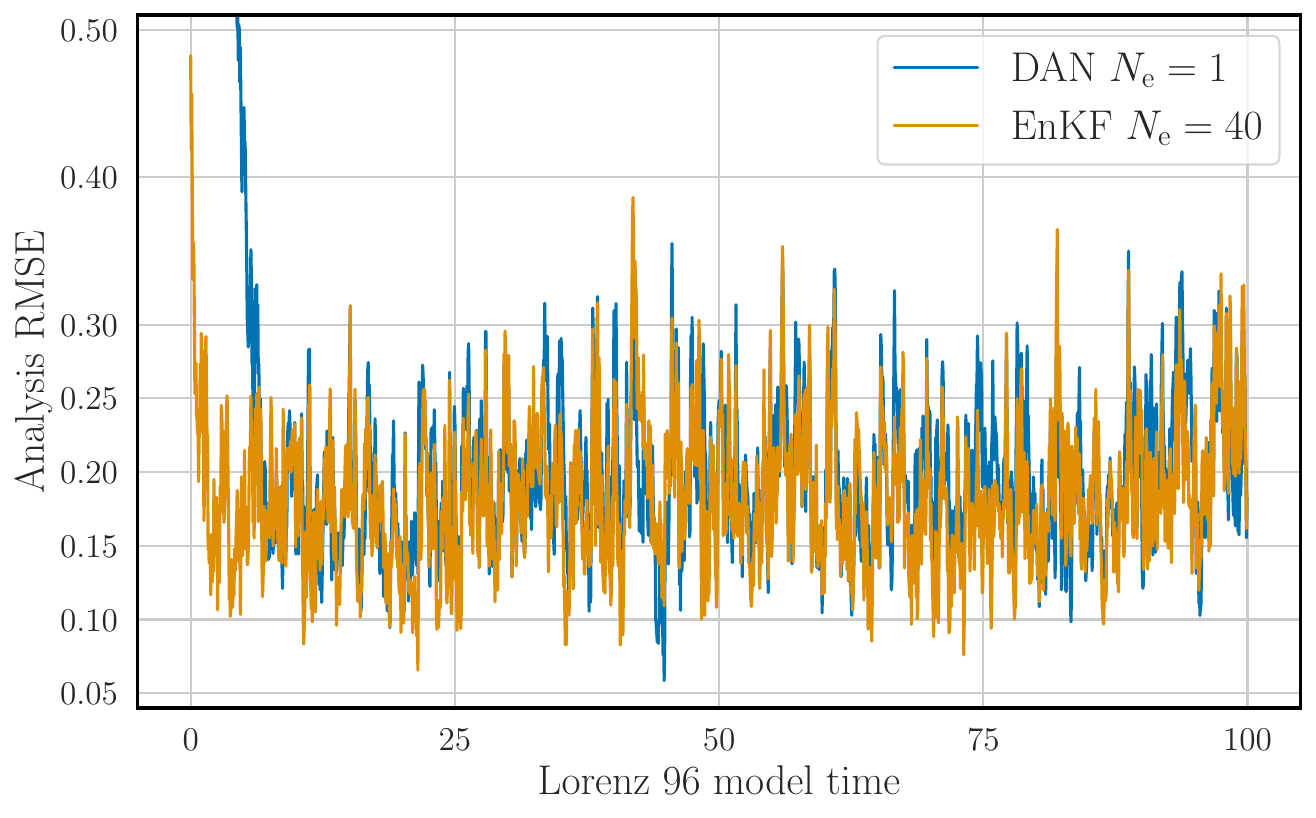}
  \caption{ \label{fig:rmse_time_series_l96} Instant RMSEs of the DAN scheme ($\Ne=1$) and of a well-tuned EnKF
    ($\Ne=40$) as a function of the L96 model time.}
\end{figure}

Figure~\ref{fig:rmse_time_series_l96} illustrates time series of instant RMSEs for the DAN scheme in the reference
configuration with $\Ne=1$, and for a well-tuned EnKF with $\Ne=40$, actually the EnKF-N. These time series include the
initial spin-up time period. Note that the EnKF starts from a perturbed true state, whereas DAN starts from a random
state on the L96 attractor.

\end{document}